\documentclass[11pt]{article}
\usepackage[utf8]{inputenc}
\usepackage[left=2.5cm, right=2.5cm, top=2.5cm]{geometry}
\usepackage{natbib} 
\usepackage{tikz} 
\usepackage{setspace} 
\usepackage{amsmath}  
\usepackage{amssymb}
\usepackage{algorithm}
\usepackage{algpseudocode}
\usepackage{booktabs} 
\usepackage{tabularx} 
\usepackage{pdflscape} 
\usepackage{mathtools} 
\usepackage{listings}
\usepackage[column=O]{cellspace} 
\usepackage{multibib}  
\newcites{APX}{Appendix References}

\usepackage{enumitem}  

\usepackage{rotating}  
\usepackage{xparse}  
\NewDocumentCommand{\rot}{O{90} O{.005em} m}{\makebox[#2][l]{\rotatebox{#1}{#3}}}

\usepackage{caption}  
\usepackage{subcaption}

\pgfdeclarelayer{bg}    
\pgfsetlayers{bg,main}  

\usepackage[normalem]{ulem}
\useunder{\uline}{\ul}{}
\usepackage[section]{placeins}  

\usepackage{ragged2e}

\usepackage{authblk}

 \usepackage{longtable}

\makeatletter
\g@addto@macro\normalsize{%
  \setlength\abovedisplayskip{8pt}
  \setlength\belowdisplayskip{8pt}
  \setlength\abovedisplayshortskip{2pt}
  \setlength\belowdisplayshortskip{2pt}
}
\makeatother

\xdefinecolor{ut-red}{RGB}{165,30,55}
\xdefinecolor{ut-gold}{RGB}{180,160,105}
\xdefinecolor{ut-dark}{RGB}{50,65,75}

\usetikzlibrary{shapes,decorations,arrows,calc,arrows.meta,fit,positioning}
\tikzset{
    -Latex,auto,node distance =1 cm and 1 cm,semithick,
    state/.style ={ellipse, draw, minimum width = 0.7 cm},
    point/.style = {circle, draw, inner sep=0.04cm,fill,node contents={}},
    bidirected/.style={Latex-Latex,dashed},
    el/.style = {inner sep=2pt, align=left, sloped}
}
\usepackage{siunitx}  
\usepackage[hidelinks]{hyperref}  

\title{Double Machine Learning meets Panel Data - Promises, Pitfalls, and Potential Solutions}
\author[]{Jonathan Fuhr}
\author[]{Dominik Papies}
\affil[]{School of Business and Economics, University of Tübingen,  Tübingen, Germany}
\date{Last edited: \today}

\begin{document}
\doublespacing

\maketitle

\begin{abstract}  
Estimating causal effect using machine learning (ML) algorithms can help to relax functional form assumptions if used within appropriate frameworks. However, most of these frameworks assume settings with cross-sectional data, whereas researchers often have access to panel data, which in traditional methods helps to deal with unobserved heterogeneity between units. In this paper, we explore how we can adapt double/debiased machine learning (DML) \citep{chernozhukov_doubledebiased_2018} for panel data in the presence of unobserved heterogeneity. This adaptation is challenging because DML's cross-fitting procedure assumes independent data and the unobserved heterogeneity is not necessarily additively separable in settings with nonlinear observed confounding. We assess the performance of several intuitively appealing estimators in a variety of simulations. While we find violations of the cross-fitting assumptions to be largely inconsequential for the accuracy of the effect estimates, many of the considered methods fail to adequately account for the presence of unobserved heterogeneity. 
However, we find that using predictive models based on the correlated random effects approach \citep{mundlak_pooling_1978} within DML leads to accurate coefficient estimates across settings, given a sample size that is large relative to the number of observed confounders. We also show that the influence of the unobserved heterogeneity on the observed confounders plays a significant role for the performance of most alternative methods.  

\end{abstract}

\clearpage
\section{Introduction}

Across multiple quantitative disciplines, recent years have seen an explosion of research on methodologies that try to use machine learning (ML) to help estimate causal effects \citep[e.g.,][]{athey_generalized_2019, chernozhukov_doubledebiased_2018}. 
These methods aim to relax assumptions in the causal estimation process by using modern ML methods to learn certain properties of the data. 
Arguably one of the most popular of these methods is the double/debiased machine learning (DML) framework by \citet{chernozhukov_doubledebiased_2018}. DML can help relax assumptions about how to adjust for observed confounders by modeling the confounding relationships with flexible ML methods. That is, in the case of a large number of potentially important confounders, or in cases where the functional forms of the confounding influences are unknown, DML uses flexible ML methods to pick the most important confounders and adjust for them flexibly. 
This framework has seen applications in a variety of disciplines \citep[e.g.,][]{felderer_using_2023, gordon_close_2022, parpouchi_association_2021}, as well as further developments and extensions to settings beyond the original ones \citep[e.g.,][]{bodory_evaluating_2022, chiang_multiway_2022, liu_doubledebiased_2021}.

At the same time, many more traditional methods from statistics and econometrics are still the default approaches for credible causal effect estimation. This is mostly because they aim to relax assumptions that are stronger than the estimation assumptions DML can relax. These methods address assumptions about causal identification, e.g., how to estimate causal effects in the presence of unobserved confounding. Examples are panel data methods, difference-in-differences, synthetic control, instrumental variables, and regression discontinuity designs \citep[e.g.,][]{cunningham_causal_2021, huntington-klein_effect_2022}. While there has been progress in using some of these methods within the DML framework \citep[see, e.g.,][]{chernozhukov_applied_2024}, very little research has explored how to adapt DML to settings with panel data, which will be the focus of our paper. 

In many applications, we observe the same units (e.g., individuals, firms, cities, etc.) repeatedly over time. This kind of data structure -- panel data -- can be very beneficial for the identification and estimation of causal effects, since it can help to eliminate any time-constant source of unobserved confounding or heterogeneity between units  \citep{wooldridge_econometric_2010}.
However, even when using panel data in this way, we could still benefit from methods that flexibly adjust for the \textit{observed} and \textit{time-varying} confounders. Hence, combining a method like DML with panel data methods could enable us to relax both assumptions about unobserved confounding and about functional forms in the observed confounding. 
For example, we might be interested in estimating the effect of price on demand for a product repeatedly sold across various stores. Our data might contain information about advertising and promotions, which act as covariates we would want to flexibly adjust for with a method such as DML. However, there could also be time-constant unobserved store characteristics influencing both the price and the demand (e.g., the management quality),  which we can potentially eliminate by exploiting the panel structure.

Unfortunately, it does not yet seem obvious how we can use DML for panel data in the presence of unobserved heterogeneity. \citet{chernozhukov_doubledebiased_2018} developed the original method and its statistical guarantees in a cross-sectional setting under the assumption of unconfoundedness. Panel data poses two major problems for DML: (1) DML uses a form of sample splitting called ``cross-fitting", which relies on i.i.d.\ data and becomes complicated if another dimension (e.g., time) is present. (2) In linear panel data models, we can use fixed effects to eliminate time-constant confounding \citep[e.g.,][]{wooldridge_econometric_2010}, but only if the parametric model is correctly specified. On the other hand, within the DML algorithm, it is not straightforward where we can similarly handle unobserved heterogeneity, especially in the settings with nonlinear confounding influences, in which the original DML excels.

In this paper, our goal is to explore the potentials and problems that using DML for panel data can pose. We state the challenges, consider different potential solutions, evaluate them in a variety of simulations, clarify and discuss the necessary assumptions, and finally give recommendations for applied researchers when using DML in panel data settings. 
Our focus is on whether different point estimators can reliably recover a known true causal effect from panel data, potentially in the presence of unobserved heterogeneity (as opposed to deriving asymptotic properties and constructing variance estimators). 

To preview our results, we find that violations of the independence assumption within the cross-fitting procedure are less consequential for the estimated coefficients than expected. However, many of the considered estimation methods struggle to remove the unobserved heterogeneity in settings with nonlinear observed confounding. For example, a seemingly natural approach of conducting DML on the time-demeaned variables (similar to fixed effects estimation) is strongly biased in settings with nonlinear confounding. We also show that the influence of the unobserved heterogeneity on the observed confounders has an impact on the accuracy of several methods. 
Nevertheless, an alternative approach that explicitly models the unobserved heterogeneity in the ML models within DML, based on the correlated random effects approach \citep{mundlak_pooling_1978}, performs well across a variety of settings. Since explicitly modeling the unobserved heterogeneity involves the introduction of additional predictors in the ML models, this approach requires the sample size being large relative to the number of observed confounders. 

The remainder of our paper proceeds as follows: Section 2 briefly reviews DML and traditional panel data methods, before discussing literature at the intersection of these research streams. Section 3 states the challenges when using DML for panel data and considers several solutions for each challenge. In Section 4, we assess these potential solutions on simulated data, generated from a variety of data-generating processes, and point out advantages and drawbacks of each method. Section 5 concludes with a discussion of the results and derives recommendations for using DML with panel data in practice.

\section{Literature overview}

Our paper contributes to the literature by drawing together research around DML and the classic econometric literature around panel data, seeking to explore how these research streams can be mutually beneficial. We therefore first review the original DML method, as well as textbook panel data methods. Then, we survey work that aims to use general ML methods with panel data, extends DML to similar settings, or directly tries to adapt DML to work with panel data.

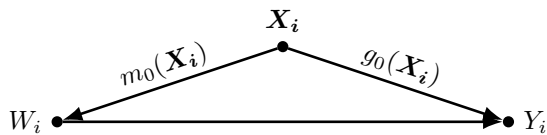
\begin{figure}[ht]
\small
\centering
\begin{tikzpicture}
       \node (w) at (0,0) [label=left:$W_i$,point, line width = 1pt];
       \node (y) at (6,0) [label=right:$Y_i$,point, line width = 1pt];
       \node (xc) at (3,1) [label=above:$\boldsymbol{X_i}$, point, line width = 1pt];
       \path[line width = 1pt] (w) edge (y);
       \path[line width = 1pt] (xc) edge node[above, el] {$g_0(\boldsymbol{X_i})$} (y);
       \path[line width = 1pt] (xc) edge node[above, el] {$m_0(\boldsymbol{X_i})$} (w);
\end{tikzpicture}
\caption{\label{fig:dagconf}Causal graph for the assumed causal structure ``unconfoundedness". $W_i$: treatment variable, $Y_i$: outcome variable, $\boldsymbol{X_i}$: observed confounding variables. The relationships between $\boldsymbol{X_i}$ and $W$ ($m_0()$), and $\boldsymbol{X_i}$ and $Y_i$ ($g_0()$), are potentially complex and nonlinear.}
\vspace{-2ex}
\end{figure}

DML \citep{chernozhukov_doubledebiased_2018} is a general estimation framework that allows using modern ML methods to flexibly adjust for observed confounding influences. Using flexible ML methods instead of a predetermined (e.g., linear) model helps to relax assumptions about variable selection and functional forms in settings where we observe all important confounders. DML does not directly address the problem of \textit{unobserved} confounding (though it can help in instrumental variables settings to make the (conditional) exogeneity assumption of instruments more credible).  
We illustrate the intuition behind the DML algorithm for a data-generating process (DGP) based on the causal graph in Figure \ref{fig:dagconf}. Here, we want to estimate the causal effect of a treatment $W_i$ on an outcome $Y_i$. However, confounders $\boldsymbol{X_i}$ influence both treatment and outcome and therefore bias the estimation if we do not adequately adjust for them. In practice, researchers often attempt to adjust for such confounders by including the corresponding variables linearly (or with another prespecified functional form) in a regression model. Yet the true way the confounders influence treatment and outcome (the functions $m_0()$ and $g_0()$) might be unknown and potentially complex. This is where DML suggests modeling these influences with flexible ML methods, which are potentially capable of learning these relationships from the data. 
For this, the DML algorithm proceeds in five steps (here, illustrated for the partially linear model): (1) Randomly split the dataset into $K$ folds, (2) hold out one fold, train two ML models on the remaining $K-1$ folds: first, predict treatment $W_i$ from confounders $\boldsymbol{X_i}$; second, predict outcome $Y_i$ from confounders $\boldsymbol{X_i}$, (3) make predictions for treatment and outcome from the estimated models, using the held out fold, and subtract them from the true values to obtain residuals, (4) use OLS to regress the outcome residual on the treatment residual and obtain the coefficient, (5) repeat steps (2)-(4) for each of the $K$ folds, and average the resulting coefficients to get the final estimate. 

\citet{chernozhukov_doubledebiased_2018} coin the term ``cross-fitting" for the technique of splitting the data, doing training and effect estimation on separate samples, and repeating the procedure for all samples. The random splitting in the cross-fitting procedure relies on the observations being independent and identically distributed (i.i.d.) \citep{chernozhukov_doubledebiased_2018}, which is one of the challenges arising for applications with clustered or panel data. We will further elaborate on this issue when we describe potential ways of combining DML with panel data.

On the other hand, a major emphasis in econometric research is finding methods that can estimate causal effects even in the presence of unobserved confounding. We can overcome one particular type of such omitted variable bias by exploiting the special structure of panel data -- observing the same units repeatedly over time \citep[e.g.,][]{wooldridge_econometric_2010}. This type of data is very prevalent in many research fields, where the units could represent individuals, firms, products, cities, etc. 
If some unobserved confounders vary only between units, but not over time for the same unit, we can use panel data estimators (e.g., the fixed effects estimator) to eliminate any such time-constant unobserved heterogeneity \citep{wooldridge_econometric_2010}. By eliminating any time-constant confounding, we can relax the assumption of ``no unobserved confounding" to ``no time-variant unobserved confounding", which is considerably more plausible in many applications. 

The fixed effects (FE) estimator works by either time-demeaning the data or by using a dummy variable regression \citep{wooldridge_introductory_2012}. In the time-demeaning approach, we subtract the time mean of each variable from the variable itself. Through this ``within-transformation", we eliminate any (even unobserved) time-constant variables, since they are identical to their time means \citep{wooldridge_introductory_2012}. In the equivalent dummy variable approach, we simply include a binary/dummy variable for each unit in the regression. While this leads to a large number of explanatory variables (which can be problematic with small samples and many units), it gives the identical effect estimates as the time-demeaning approach in two-dimensional panels \citep{wooldridge_introductory_2012}. 
Transferring either approach to the DML framework is not trivial. One challenge is the decision at which step of the DML algorithm the time-demeaning or the inclusion of dummy variables should occur. The first option is ``early", meaning we demean all variables before training the two ML models or include the dummy variables already in the training of the ML models. The alternative is ``late", meaning that after computing the residuals in step (3), we time-demean the residuals before running the residual-on-residual regression, or we include the dummy variables directly in the final regression without demeaning first. We will discuss potential benefits and drawbacks of each of these approaches in Sections \ref{sec:methods} and \ref{sec:sims}, but also present another option that mostly overcomes these challenges.

One lesser known alternative to the fixed effects estimator is the correlated random effects (CRE) approach \citep[e.g.,][]{wooldridge_econometric_2010}. Whereas on the one hand, a classical random effects approach assumes no correlation between the unobserved heterogeneity and the covariates, and on the other hand, the fixed effects approach does not restrict this relationship at all, the CRE framework unifies these approaches by explicitly modeling this correlation \citep{wooldridge_introductory_2012}.
The initial suggestion by \citet{mundlak_pooling_1978} was to let the unobserved heterogeneity be correlated with the \textit{average} level of the covariates over time, which amounts to including the time means of each covariate in a pooled OLS or random effects regression (in addition to the original variables which vary by both time and unit). Later approaches by \citet{chamberlain_multivariate_1982, chamberlain_chapter_1984} instead model the unobserved heterogeneity with a linear model of the covariates' time history. The attractive feature of these CRE approaches in our setting is that at least in the linear case, they give coefficient estimates identical to the FE approach \citep{wooldridge_introductory_2012}, while more explicitly modeling the unobserved heterogeneity instead of removing it, which could prove useful for the ML prediction steps.

\def\arraystretch{1.4}
\newcolumntype{L}[1]{>{\raggedright\let\newline\\\arraybackslash\hspace{0pt}}m{#1}}
\newcolumntype{C}[1]{>{\centering\let\newline\\\arraybackslash\hspace{0pt}}m{#1}}
\newcolumntype{R}[1]{>{\raggedleft\let\newline\\\arraybackslash\hspace{0pt}}m{#1}}

\begin{table}[ht!]
    \caption{Methods at the intersection of DML and panel data}
\small
    \centering
    \resizebox{\columnwidth}{!}{
    \begin{tabular}{L{.2\textwidth} L{.15\textwidth}  L{.15\textwidth} L{.2\textwidth} L{.15\textwidth} L{.15\textwidth}}
    \toprule
        \textbf{Method} & \textbf{Reference} & \textbf{ML algorithms}  &  \textbf{Splitting procedure} & \textbf{Made for panel data} & \textbf{Considers unobs. het.} \\ \midrule 
         DML & \citet{chernozhukov_doubledebiased_2018} & Any  & Standard cross-fitting & No & No  \\
         Fixed effects / CRE & \citet{wooldridge_econometric_2010} & n.a. & None & Yes & Yes \\
         Cluster-Lasso & \citet{belloni_inference_2016} & Only lasso & None & Yes & Yes (early demeaning) \\
         DML for Diff-in-diff & \citet{chang_doubledebiased_2020} & Any & Standard cross-fitting  & Not really (only 2 periods) & No  \\
         Multiway DML & \citet{chiang_multiway_2022} & Any & Multiway cross-fitting & No (but clustered data) & No \\
        (Debiased) orthogonal lasso & \citet{semenova_inference_2023} & Only lasso & Neighbors-left-out cross-fitting & Yes & Yes (sparse FEs) \\ 
        Extensions of within-group, first-difference, and CRE & \citet{clarke_double_2023} & Lasso, regression trees, random forests (any) & Split by unit in cross-fitting & Yes & Yes \\ \bottomrule
    \end{tabular}
    }
    \label{tab:lit_dml_panel}
\end{table}

In Table \ref{tab:lit_dml_panel}, we summarise some features of standard DML, the fixed-effects estimator, and further developments related to DML that have explored panel data or similar settings. The first two rows show the previously described original DML framework and the traditional fixed effects estimator. The DML framework uses cross-fitting and allows for usage of any well-performing ML algorithm, but assumes a cross-sectional setting and does not consider any unobserved heterogeneity. On the other hand, the traditional fixed effects estimator and the correlated random effects approach can handle time-constant unobserved heterogeneity in panel data, whereas they do not use any ML algorithms (and no sample splitting) to make adjustment for time-variant variables more flexible. 
We use these two methods as a starting point and explore how the literature up to the time of writing this article (April 2024) has attempted to harmonize ML-based effect estimation with panel data settings.  

First, \citet{belloni_inference_2016} published a method for effect estimation in panel data settings with potentially high-dimensional time-varying confounders, even before the publication of DML in \citet{chernozhukov_doubledebiased_2018}. They treat the unobserved heterogeneity as fixed effects and remove them through demeaning in a first step. Then, they use a variant of lasso that accounts for the clustering of units (Cluster-Lasso) to select important predictors in the treatment and in the outcome model, and use the selected variables as controls in a final OLS regression of the (demeaned) outcome on the treatment \citep{belloni_inference_2016}. While lasso can learn the correct model under certain sparsity assumptions even without sample splitting, researchers must manually decide which variable transformations to include in the algorithm, since the authors' method does not facilitate the use of more flexible ML methods such as random forests or neural networks \citep{chernozhukov_doubledebiased_2018}. 

Second, \citet{chang_doubledebiased_2020} extends the DML framework to the semiparametric difference-in-differences estimator, for situations where the parallel trends assumption only holds after adjusting flexibly for controls. While their setting does have a time dimension, they only use it to derive a time indicator for post-treatment, which indicates whether an observation comes from before or after reception of the (binary) treatment. They then derive a DML estimator for the average treatment effect on the treated (ATT), which is very similar to the ATT estimator in \citet{chernozhukov_doubledebiased_2018}, but uses the difference in observed outcomes instead of the outcome itself \citep[see also][]{chernozhukov_applied_2024}. 
Moreover, they only consider time-constant confounders and therefore do not deal with unobserved unit-level heterogeneity. This setting is thus closer to the classical cross-sectional setting and does not encounter the problems we face when dealing with panel data in DML.

The third method also does not directly consider panel data but instead extends DML to multiway clustered data \citep{chiang_multiway_2022}. This setting is related to panel data in that the data is also double indexed, but instead of a time dimension there are two distinct unit/cluster dimensions (e.g., markets and products). To account for the clustered structure, \citet{chiang_multiway_2022} develop a novel multiway ($K^2$-fold) cross-fitting procedure that enables DML with any ML algorithm, even if the data is not i.i.d., but clustered along multiple dimensions. However, the authors' main goal is \textit{valid inference} in the multiway clustered setting, they do not address identification by accounting for cluster-specific unobserved heterogeneity. Also, their method does not need to consider the unique challenges of the time dimension in cross-fitting.

Fourth, \citet{semenova_inference_2023} develop methods for conditional average treatment effects in high-dimensional dynamic panel data settings. They build on the correlated random effects approach \citep{mundlak_pooling_1978} and assume approximate sparse fixed effects (i.e., only few fixed effects are important/nonzero). Their ``neighbors-left-out" splitting procedure adapts cross-fitting to support weakly dependent (panel) data. However, the authors rely on a version of lasso as the only possible ML algorithm for panel data and only hint at the potential of other ML methods in such a setting. 

Finally, in an independent development parallel to ours, \citet{clarke_double_2023} develop DML estimators for panel data that account for unobserved individual heterogeneity in the partially linear model. They implement their approaches with lasso, regression trees, and random forests, but can in principle incorporate any ML algorithm. For the cross-fitting, they split the data by unit, such that the full time series of each unit ends up in the same fold. 
Their first approach is an ``approximation approach", where they first transform the raw data to remove the unobserved heterogeneity, and then use DML on the transformed data. The alternative approaches integrate the correlated random effects model by \citet{mundlak_pooling_1978} in a DML framework. In simulations with nonlinear and discontinuous settings, \citet{clarke_double_2023} find superior performance of the latter methods built on CRE compared to standard linear methods or the approximation approach. However, they also demonstrate that DML with tree-based ML algorithms does not lead to valid inference in their settings. 

Our study contributes to this literature in general and \citet{clarke_double_2023} specifically in four ways: (1) In addition to estimators similar to those in \citet{clarke_double_2023}, we consider further intuitively appealing approaches for dealing with unobserved heterogeneity within DML; (2) we assess these approaches in a substantially wider variety of simulation settings that differ with respect to the true DGP, revealing, e.g., the particular sensitivity of the most promising approach to an increasing number of observed confounders; (3) we explicitly demonstrate and discuss the consequences of DGPs where the unobserved heterogeneity also influences the observed confounding, which is likely to occur in many applications, and (4) we investigate the impact of various distinct splitting strategies in the cross-fitting procedure on the estimated effects.

\section{Possible methods for DML with panel data} \label{sec:methods}

In this section, we provide a detailed description of the problems that emerge when using DML for settings with panel data, and consider adaptations of DML to address these issues. We identify two problems: (1) the sample-splitting/cross-fitting procedure of DML with dependent data, and (2) accounting for potential unobserved heterogeneity in DML. For both problems, we explore a variety of possible solution approaches in the following.

\subsection{Different splitting strategies for DML with panel data} \label{sec:meth_split}

One essential component of the original DML framework \citep{chernozhukov_doubledebiased_2018} is the cross-fitting procedure. This kind of sample splitting is important to remove overfitting bias that can arise if we use the same observations for training the ML models and estimating the effects \citep{chernozhukov_doubledebiased_2018}. To avoid this overfitting in DML, we train the ML methods on a part of the data, but make predictions and estimate the effects on another part that we did not use for training. In general, sample splitting leads to less efficient estimation, since we use only part of the data for training and estimation, respectively. Nevertheless, DML regains full efficiency by switching the roles of the training and estimation sample and averaging the resulting estimates (``cross-fitting") \citep{chernozhukov_doubledebiased_2018}. 

However, cross-fitting relies on the assumption that the observations are independent and identically distributed (i.i.d.) \citep{chernozhukov_doubledebiased_2018}. As soon as we enter settings with panel data, this assumption is violated, because data points are dependent over time (i.e., serial correlation/autocorrelation) and/or within a cluster, and can end up in the same or in different samples when randomly splitting the data \citep{chiang_multiway_2022}. As a consequence, there is a danger of overfitting the ML models to the hold-out data. While this certainly makes consistency statements, asymptotic analysis, and valid confidence intervals difficult \citep[e.g.,][]{wooldridge_econometric_2010}, it is unclear how severe the practical consequences for the estimated effect coefficients really are. In our study, we will assess how different splitting strategies affect the finite-sample performance of different DML estimators. We leave the analysis of asymptotic properties and the construction of valid confidence intervals to future research.

When implementing cross-fitting for panel data, there are a variety of options for how to split the sample (Table \ref{tab:split_DML_approaches}). 
First, we could do the standard random split as in the original DML algorithm for the cross-sectional setting, ignoring the special panel structure. This can potentially lead to the overfitting bias previously described.

\def\arraystretch{1.3}

\begin{table}[ht]
\centering
\small
\caption{Different approaches to sample splitting when using DML with panel or clustered data}
\label{tab:split_DML_approaches}
\resizebox{\columnwidth}{!}{
\begin{tabular}{p{.2\columnwidth} p{.4\columnwidth} p{.4\columnwidth}}
\toprule
\textbf{Splitting strategy} & \textbf{Description}  & \textbf{Problems} \\ \midrule
Random              & Random splitting as in classic DML, ignoring panel structure & Ignores panel structure and dependency within units/across time \\
By unit              & All observations of the same \textit{unit} end up in the same fold, otherwise random & ML methods cannot predict unit-specific effects in the hold-out fold; time dependence potentially still present \\
By time (period)     & All observations of the same \textit{period} end up in the same fold, otherwise random & Time dependence of period-adjacent observations; 
cluster dependence potentially still present \\
By time (folds)      & $T/K$ \textit{adjacent} periods end up in the same fold  & Time dependence at the splitting point of adjacent folds; 
cluster dependence potentially still present  \\
By time (neighbors-left-out)   & $T/K$ \textit{adjacent} periods end up in the same fold, folds adjacent to the prediction fold are excluded from training & 
Cluster dependence potentially still present \\
 \bottomrule
\end{tabular}
}
\end{table}


Secondly, we could split the data by the unit dimension and ensure that observations of each unit (index $i$) end up in only the training or the estimation sample at any point (as done in, e.g., \citet{clarke_double_2023}). While this ensures independence between training and estimation fold in the unit dimension, it becomes problematic if we expect the ML methods to also model unit-specific effects \citep{semenova_inference_2023}. That is, we cannot model unit-specific unobserved heterogeneity by, e.g., including unit dummies as predictors in the ML algorithm, since the units in the hold-out sample (used for prediction and estimation of effects) are not present in the training sample. Thus, when splitting by unit, we cannot remove the unobserved heterogeneity with such an approach. 

Alternatively to splitting by unit, we could split along the other dimension, i.e., by time. Here we consider three different options. These options have in common that they do not address the potential cluster dependence along the unit dimension. 
First, we can ensure that all observations of the same period (index $t$) end up in the same sample. However, this does not help in the case of serial correlation, since the adjacent observations in the time dimension can still end up in the other sample and thus induce a dependence. 

The fourth splitting strategy improves upon the previous one by splitting the time-ordered data into $K$ folds. That is, observations with indices from $t = 1$ to $t = T/K$ end up in the first fold, from $t = T/K + 1$ to $t = 2T/K$ in the second fold, etc. This procedure reduces the time dependence to only be substantial at the splitting point of adjacent folds: The last observation of the first fold and the first observation of the second fold might be correlated, but the further one moves from the splitting point, the smaller the correlation gets. 

To also eliminate the correlations around the splitting points, \citet{semenova_inference_2023} propose a strategy they call ``neighbors-left-out cross-fitting". They suggest dividing the data into a relatively large number of time-adjacent folds ($K \ge 10$), of which we hold out not only the fold used for prediction and estimation, but also the folds in its immediate neighborhood. By this, the training and the estimation samples should be approximately independent, even in weakly dependent time series or panel data. See Appendix \ref{apx:split_illus} for a graphical illustration of the final two cross-fitting approaches.

\subsection{Accounting for unobserved heterogeneity in DML}

The second challenge when using DML with panel data is accounting for unobserved heterogeneity. One of the main motivations for using panel data in the first place is to account for unobserved influences that only vary along one dimension and are constant in the other \citep[e.g.,][Chapter 10]{wooldridge_econometric_2010}. This can, for example, be a unit-specific but time-constant unobserved variable such as $U_i$ in Figure \ref{fig:dag_panel} or in the partially linear model of Equations \ref{eq:meth_out} and \ref{eq:meth_treat}. If this variable influences both treatment and outcome, it acts as an unobserved confounder and leads to an omitted variables bias if we employ standard cross-sectional methods \citep{wooldridge_econometric_2010}. 
We will later consider three different DGPs that differ with respect to which variables the unobserved heterogeneity influences (see Figure \ref{fig:dag_panel}). However, since we typically do not know the true DGP in practice, the ideal method should perform well in each of these settings. 
\begin{figure}[ht]
\footnotesize
\centering
\begin{subfigure}[t]{0.32\textwidth}
\centering
    \begin{tikzpicture}
        \node (w) at (0,0) [label=left:$W_{it}$,point, minimum size = 1pt];
        \node (y) at (3,0) [label=right:$Y_{it}$,point, minimum size = 1pt];
        \node (xc) at (1.5,1.5) [label=above:$\boldsymbol{X_{it}}$, point, minimum size = 1pt];
        \node (u) at (-1,1.5) [fill = white, label=above:$U_i$,point, minimum size = 1pt];
        \path (w) edge (y);
        \path (xc) edge node[above, el] {\scriptsize$g_0(\boldsymbol{X_{it}})$} (y);
        \path (xc) edge node[above, el] {\scriptsize$m_0(\boldsymbol{X_{it}})$} (w);
    \end{tikzpicture}
\caption{}
\label{fig:dag_noU}
\end{subfigure}
\begin{subfigure}[t]{0.32\textwidth}
\centering
    \begin{tikzpicture}
        \node (w) at (0,0) [label=left:$W_{it}$,point, minimum size = 1pt];
        \node (y) at (3,0) [label=right:$Y_{it}$,point, minimum size = 1pt];
        \node (xc) at (1.5,1.5) [label=above:$\boldsymbol{X_{it}}$, point, minimum size = 1pt];
        \node (u) at (-1,1.5) [fill = white, label=above:$U_i$,point, minimum size = 1pt];
        \path (w) edge (y);
        \path (xc) edge node[above, el] {\scriptsize$g_0(\boldsymbol{X_{it}})$} (y);
        \path (xc) edge node[above, el] {\scriptsize$m_0(\boldsymbol{X_{it}})$} (w);
        \path[loosely dashed] (u) edge (w);
        \path[loosely dashed] (u) edge (y);
    \end{tikzpicture}
\caption{}
\label{fig:dag_noX}
\end{subfigure}
\begin{subfigure}[t]{0.32\textwidth}
\centering
    \begin{tikzpicture}
        \node (w) at (0,0) [label=left:$W_{it}$,point, minimum size = 1pt];
        \node (y) at (3,0) [label=right:$Y_{it}$,point, minimum size = 1pt];
        \node (xc) at (1.5,1.5) [label=above:$\boldsymbol{X_{it}}$, point, minimum size = 1pt];
        \node (u) at (-1,1.5) [fill = white, label=above:$U_i$,point, minimum size = 1pt];
        \path (w) edge (y);
        \path (xc) edge node[above, el] {\scriptsize$g_0(\boldsymbol{X_{it}})$} (y);
        \path (xc) edge node[above, el] {\scriptsize$m_0(\boldsymbol{X_{it}})$} (w);
        \path[loosely dashed] (u) edge (xc);
        \path[loosely dashed] (u) edge (w);
        \path[loosely dashed] (u) edge (y);
    \end{tikzpicture}
\caption{}
\label{fig:dag_X}
\end{subfigure}
    \caption{\label{fig:dag_panel}Possible DGPs for panel data settings. $W_{it}$ (treatment), $Y_{it}$ (outcome), and $X_{it}$ (observed confounders) vary across both units and time. $U_{i}$ is unobserved unit-specific and time-constant heterogeneity. We consider three causal structures: \textbf{(A)} $U_i$ does not influence any other variables (or does not exist), \textbf{(B)}  $U_i$ only influences $W_{it}$ and $Y_{it}$, \textbf{(C)}  $U_i$ additionally influences $X_{it}$.}
\end{figure}
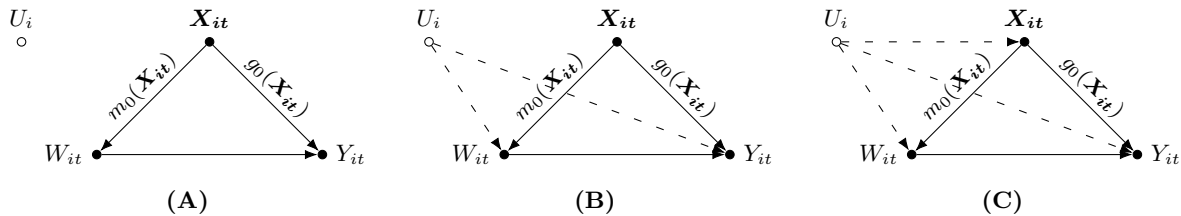
\begin{gather} 
Y_{it} = \alpha_1 + \beta W_{it} + \gamma g_0(\boldsymbol{X_{it}}) + \delta U_i + \mu_{it} \label{eq:meth_out} \\
W_{it} = \alpha_2 +  \gamma m_0(\boldsymbol{X_{it}}) + \delta U_i + \eta_{it}  \label{eq:meth_treat}
\end{gather}
In traditional econometrics, the most common way of dealing with unobserved heterogeneity in panel data is fixed effects estimation \citep[e.g.,][Chapter 10]{wooldridge_econometric_2010}. If there is unobserved confounding which varies only along one cluster dimension, using fixed effects can eliminate its biasing influence (\citealp{wooldridge_introductory_2012}, Chapter 14). Traditionally, we implement fixed effects estimation by time-demeaning all variables and running OLS on the transformed variables. Since the unobserved heterogeneity is fixed over time, it disappears from the demeaned equation \citep{wooldridge_introductory_2012}. Alternatively, we can run an OLS regression where we include a dummy/binary variable for each unit, which leads to the exact same results as fixed effects estimation in the standard setting \citep{wooldridge_introductory_2012}. 

However, if we want to use DML to flexibly adjust for observed confounding in the presence of unobserved heterogeneity, it is not obvious where and how in the algorithm we can consider fixed effects (or alternative ways of accounting for unobserved heterogeneity). Should we demean the variables before the ML predictions or demean the residuals afterwards? Can we consider the fixed effects directly within the ML step? How? 
In the following, we discuss several conceivable approaches for this setting (Table \ref{tab:FE_DML_approaches}).

First, we could ignore the potential for unobserved cluster-level confounding and simply use the original DML algorithm on the pooled data (``Pooled DML"). This is only appropriate if there is no unobserved heterogeneity; if this assumption is violated, the estimates will be biased. 

\def\arraystretch{1.3}

\begin{table}[ht]
\small
\centering
\caption{Different approaches to fixed effects within DML}
\label{tab:FE_DML_approaches}
\resizebox{\columnwidth}{!}{
\begin{tabular}{O{p{.2\textwidth}} O{p{.8\textwidth}}}
\toprule
\textbf{Approach}  & \textbf{Description} \\ \midrule
Pooled DML  & Ignores the panel dimension and unobserved heterogeneity, does standard DML on full data \\
Early demeaning    &  First demeans all variables, then runs standard DML on the demeaned data \\
Late demeaning   &  First runs steps (1)-(3) of DML on original data, then demeans the residuals before running OLS \\
DML with dummies   &  Runs standard DML, but adds a dummy for each unit in the prediction steps \\
DML with CRE    &  Runs standard DML, but adds time means for treatment and confounders in the prediction steps as in correlated random effects  \\
 \bottomrule
\end{tabular}
}
\end{table}

The second and third approach are inspired by the traditional fixed effects estimation and deal with the unobserved heterogeneity by introducing an additional step in the DML algorithm. In what we call ``early demeaning",  we try to eliminate the unobserved heterogeneity \textit{before} DML. That is, we demean all variables first, then perform DML on the demeaned variables. If we fully eliminate the unobserved heterogeneity by this fixed effects transformation, the prediction tasks in DML should be easier, since the ML methods only need to model observed variation. The ``within-transformed" outcome and treatment equations for the ML training are
\begin{equation} \label{eq:early}
\begin{split}
y_{it} - \bar{y_i} &= \gamma g_0(x_{it} - \bar{x_i}) + \mu_{it} - \bar{\mu_i}   \\
w_{it} - \bar{w_i} &= \delta m_0(x_{it} - \bar{x_i}) + \eta_{it} - \bar{\eta_i},
\end{split}
\end{equation}
where $\bar{w_i}$ is the mean treatment over time ($\frac{1}{T}\sum_{t=1}^T w_{it}$). This early demeaning is only appropriate if the time mean is additively separable in the true DGP, which holds in the linear case, but not necessarily if $g_0()$ or $m_0()$ are nonlinear, as we will show below. This approach is similar to the ``approximate approach" in \citet{clarke_double_2023}.

In the third approach, we deal with the fixed effects \textit{later} in the DML algorithm (``late demeaning"). Here, we run steps (1)-(3) of DML first (ML and residualization), before we demean the residuals in the final DML step and regress the demeaned outcome residual on the demeaned treatment residual: 
\begin{equation} \label{eq:late}
\begin{split}
v^y_{it} - \bar{v^y_i} &= v^w_{it} - \bar{v^w_i} + \epsilon_{it} - \bar{\epsilon_i} .
\end{split}
\end{equation}
This late demeaning increases the difficulty for the ML prediction tasks early in the algorithm, since the unobserved heterogeneity is still present and acts as additional noise. However, if the ML method still manages to successfully model the observed confounding, the remaining unobserved heterogeneity may be additively separable afterwards. 

In contrast to the previous two approaches, the final two strategies try to directly model the unobserved heterogeneity within the two ML models in DML. Instead of an additional step in the algorithm, these approaches change the predictors/features supplied to the ML methods.
Approach four (``DML with dummies") includes unit dummies in the prediction steps of DML, inspired by the equivalence of the dummy variable regression to fixed effects estimation in the standard linear setting \citep{wooldridge_introductory_2012}. That is, the outcome and treatment equations for the ML training are
\begin{equation} \label{eq:dummies}
\begin{split}
y_{it} &= \gamma g_0(x_{it}, z1_i, ..., zN_i) + \mu_{it}  \\ 
w_{it} &= \delta m_0(x_{it}, z1_i, ..., zN_i) + \eta_{it},
\end{split}
\end{equation}
where $z1_i$ to $zN_i$ are dummy variables for each unit, which are 1 if the observation belongs to that unit and 0 if it does not.  This should work well in settings where there are relatively few units, or in settings with high-dimensional fixed effects (i.e., many units), if we can assume that only few of these are important (``sparse" fixed effects). This sparsity assumption is common in the literature, but often not realistic in settings with unit-specific unobserved heterogeneity \citep[e.g.,][]{belloni_inference_2016}. If the assumption is violated, the prediction tasks will likely become too complex if the number of observations is not significantly larger than the number of relevant fixed effects.

The final approach (``DML with CRE") avoids this sparsity assumption about the fixed effects by explicitly modeling the relationship between the unobserved heterogeneity and the covariates as proposed in the correlated random effects approach \citep[e.g.,][]{mundlak_pooling_1978}. \citet{chernozhukov_automatic_2022} mention a similar DML approach for panel data in the application of their novel automatic debiased ML framework, and \citet{clarke_double_2023} also introduce a related CRE estimator. In the context of DML, using Mundlak-type correlated random effects amounts to including the treatment and covariate time means in addition to the time-varying covariates into the ML predictions for both the outcome and the treatment: 
\begin{equation} \label{eq:dlm_cre}
\begin{split}
y_{it} &= \gamma g_0(x_{it}, \bar{x_i}, \bar{w_i}) + \mu_{it} \\
w_{it} &= \delta m_0(x_{it}, \bar{x_i}, \bar{w_i}) + \eta_{it}.
\end{split}
\end{equation}
This way, we can model the time-constant unobserved heterogeneity without introducing a large number of additional variables. However, if $J$ is the number of covariates, the number of predictors in the ML models using the CRE approach increases by a factor of $2 * J$, since we introduce the time mean for each additional variable as well. By comparison, all other approaches only scale by the factor $J$.

\section{Simulations} \label{sec:sims}

We now explore how the suggested methods perform on simulated data. First, we give an overview of how we implemented the considered DML methods and introduce alternative traditional statistical methods as benchmarks. Then, we describe our baseline DGPs, which we subsequently use to compare different cross-fitting techniques, as well as different estimation methods. Finally, we extend our baseline simulations and change different characteristics of the DGPs to investigate the methods' sensitivity to these modified situations.\footnote{All code for method implementation, data generation, and estimation is available on OSF: \url{https://osf.io/8skxu/?view_only=1d56c1e412084ee399cd9a3fdcb39c02}.}

\subsection{Method implementations}

In addition to the various DML approaches described in the previous section, we also implemented several traditional statistical methods as comparison baselines (Table \ref{tab:methods_impl}).
The first method is a naive simple OLS regression, where we regress the outcome on the treatment but adjust neither for observed covariates nor for unobserved heterogeneity. 
As the second method, we also use an OLS regression of the outcome on the treatment, but now include all observed covariates linearly, i.e., pooled OLS (POLS) \citep{wooldridge_econometric_2010}. This method will be biased and inconsistent if there is any unobserved heterogeneity and/or if the confounding influence is not linear.
Thirdly, we use the standard fixed effects estimator, where we adjust for all covariates linearly and try to adjust for the unobserved heterogeneity by including fixed effects. 

Approaches 4-8 are the various DML implementations for panel data discussed in Table \ref{tab:FE_DML_approaches}. As the predictive ML algorithm in step (3) of DML, we use boosted trees as implemented in XGBoost.
In our implementation, we use default values for the learning rate eta (0.3) and the maximum tree depth (6), and employ early stopping if the validation set performance does not improve for 10 rounds. We tune the optimal maximum number of boosting iterations by choosing from up to 200 rounds with 5-fold cross-validation.  We experimented with using other flexible ML methods within DML, which lead to very similar results. We choose XGBoost as representative of flexible ML algorithms due to its strong performance within DML in cross-sectional settings, as well as its computational efficiency compared to other flexible methods \citep{chen_xgboost_2023}. 
One noticeable feature of the ``DML (dummies)" method is that generating unit dummies potentially leads to a very large number of variables. In settings with many units, this not only complicates the prediction task, but also substantially increases the computation time for this approach compared to alternatives. In Appendix \ref{sec:comp_eff}, we show timing results for various combinations of numbers of units and periods. In our baseline simulation with 500 units and 10 periods, computing DML with dummies for one dataset takes about 330 seconds, whereas the second slowest method (DML with CRE) is computed within less than 8 seconds. 

Finally, we include an infeasible ``oracle FE" method as an additional benchmark. Here we use the standard fixed effects framework, but always in combination with a parametric model that uses the correct (in practice unknown) functional form of the covariates. This method indicates whether researchers could in theory estimate the true effect from the data if they knew the true functional form of the confounding, i.e., specified the correct parametric model.

\def\arraystretch{1.3}

\begin{table}[ht]
\centering
\small
\caption{Description of implemented methods}
\label{tab:methods_impl}
\resizebox{\columnwidth}{!}{
\begin{tabular}{p{.2\textwidth} p{.8\textwidth}}
\toprule
\textbf{Label} & \textbf{Description} \\ \midrule
Simple OLS & Linear regression ignoring all covariates and unobserved heterogeneity \\
POLS &  Pooled OLS: Linear regression with all covariates but not dealing with unobserved heterogeneity \\
Fixed effects & Linear regression with all covariates and accounting for unobserved heterogeneity with fixed effects \\
PDML & Pooled DML: ignoring panel dimension and unobserved heterogeneity, using XGBoost as predictive algorithm \\
DML (early FE) &  Early demeaning: using standard DML with XGBoost as predictive algorithm on demeaned data \\
DML (late FE) &  Late demeaning: running steps (1)-(3) of standard DML with XGBoost as predictive algorithm first, then demeaning residuals in step (4) before OLS \\
DML (dummies) &  Standard DML with XGBoost as predictive algorithm, but adding a dummy for each unit in the prediction models \\
DML (CRE) &  DML with correlated random effects, adding time means for treatment and confounders in the prediction models  \\
Oracle FE & Standard fixed effects estimation, always knowing the true form of the confounding (infeasible) \\
 \bottomrule
\end{tabular}
}
\end{table}

\subsection{Baseline data generation}

For all simulations, we follow one of the causal graphs in Figure \ref{fig:dag_panel}. We are interested in the effect of a treatment variable $W_{it}$ (e.g., price) on an outcome variable $Y_{it}$ (e.g., demand), so we need to adjust for the observed confounder(s) $X_{it}$ (e.g., advertising), which influence both treatment and outcome. Observed confounders, treatment and outcome can vary across multiple dimensions (e.g., both unit and time) and are therefore double-indexed. Furthermore, there can exist some form of unobserved heterogeneity (e.g., store characteristics such as management quality) between different units. We model this as an unobserved variable $U_i$, which varies across only one dimension (here, the unit dimension). In our baseline simulation settings, we differentiate between three causal structures that differ in how $U_i$ influences the other variables: (A) $U_i$ does not exist, or at least exerts no influence on the other variables. In this setting, we have no unobserved heterogeneity, hence flexibly adjusting for $X_{it}$ should suffice. (B) $U_i$ does exist, but influences only the treatment and the outcome directly, not the observed confounders $X_{it}$. (C) In addition to treatment and outcome, $U_i$ also influences the observed confounders $X_{it}$ and thus impacts treatment and outcome via multiple pathways. In our example, we consider structure (C) to be the most plausible, since the unobserved management quality certainly also influences decisions on advertising and promotions.

In the baseline simulation, we only consider a single variable for each of $U_i$, $X_{it}$, $W_{it}$, and $Y_{it}$. 
In all simulations, we draw single exogenous variables from a standard normal distribution ($N(0, 1)$). If the variable is clustered and time-constant like $U_i$, we draw the values on the unit-level and replicate the same value across all time periods for a given unit.  
We generate all other variables according to Equations \ref{eq:sim_conf}-\ref{eq:sim_out}, where the inclusion of $U_i$ in each equation depends on whether we simulate the causal structure (A), (B), or (C). Intercepts and noise terms follow a standard normal distribution ($\alpha, \epsilon_{it}, \eta_{it}, \mu_{it} \sim N(0,1)$). 
$g_0()$ and $m_0()$ indicate the functional form of the \textit{observed} confounding. In our baseline simulations, we use either a linear functional form ($g_0(X_{it}) = m_0(X_{it}) = X_{it}$), for which linear methods are appropriate, or a nonlinear u-shaped functional form ($g_0(X_{it}) = m_0(X_{it}) = {X_{it}}^2$), for which linear methods are misspecified, but for which flexible ML methods might be capable of learning the correct functional form.\footnote{We experimented with other nonlinear functional forms (e.g., a discontinuous step function), which led to very similar results.}
We draw the coefficients for the influence of the observed confounders ($\gamma$) and for the unobserved confounders ($\delta$) for each simulation from a standard normal distribution ($\gamma, \delta \sim N(0,1)$). 
We set the true causal effect of interest $\beta$ to one ($\beta = 1$). 
The main goal of the simulations is to investigate how well different methods can recover this coefficient across various settings. We specify the number of units $N$ and the number of periods $T$ in each of the following result sections.
\begin{gather}
X_{it} = \alpha_0 + \delta U_i + \epsilon_{it} \label{eq:sim_conf} \\ 
W_{it} = \alpha_1 +  \gamma g_0(X_{it}) + \delta U_i + \eta_{it} \label{eq:sim_treat} \\
Y_{it} = \alpha_2 + \beta W_{it} + \gamma m_0(X_{it}) + \delta U_i + \mu_{it}  \label{eq:sim_out}
\end{gather}

\subsection{Comparison of cross-fitting techniques} \label{comp_cf}

Before we compare the performance of the considered estimators, we investigate how the different cross-fitting techniques introduced in Section \ref{sec:meth_split} affect the coefficient estimates. For this, we simulate data according to the baseline DGP with the most complex causal structure (C) and u-shaped confounding influences. The dataset is a balanced panel with $N = 100$ units across $T = 50$ periods. 
We choose this relatively large number of periods to still have multiple periods in the hold-out fold when splitting the data into time-adjacent folds. 
We violate the original cross-fitting assumption of i.i.d.\ data both with the presence of unobserved unit heterogeneity and with a relatively large degree of autocorrelation. We introduce autocorrelation by changing the DGP of the outcome model (Equation \ref{eq:sim_out}) to include serially correlated errors: Now, we simulate $\mu_{it}$ as a weakly dependent time series according to an AR(1) model, i.e., $\mu_{it} = \rho\mu_{it-1} + e_{it}$, with the ar-coefficient $\rho = 0.9$.
If the cross-fitting procedure affects the coefficient estimates, it should be especially visible when violating the independence assumption. For each considered DML estimator, we implement all of the cross-fitting techniques from Table \ref{tab:split_DML_approaches}.

\begin{figure}[ht]
    \centering
    \includegraphics[width=\textwidth]{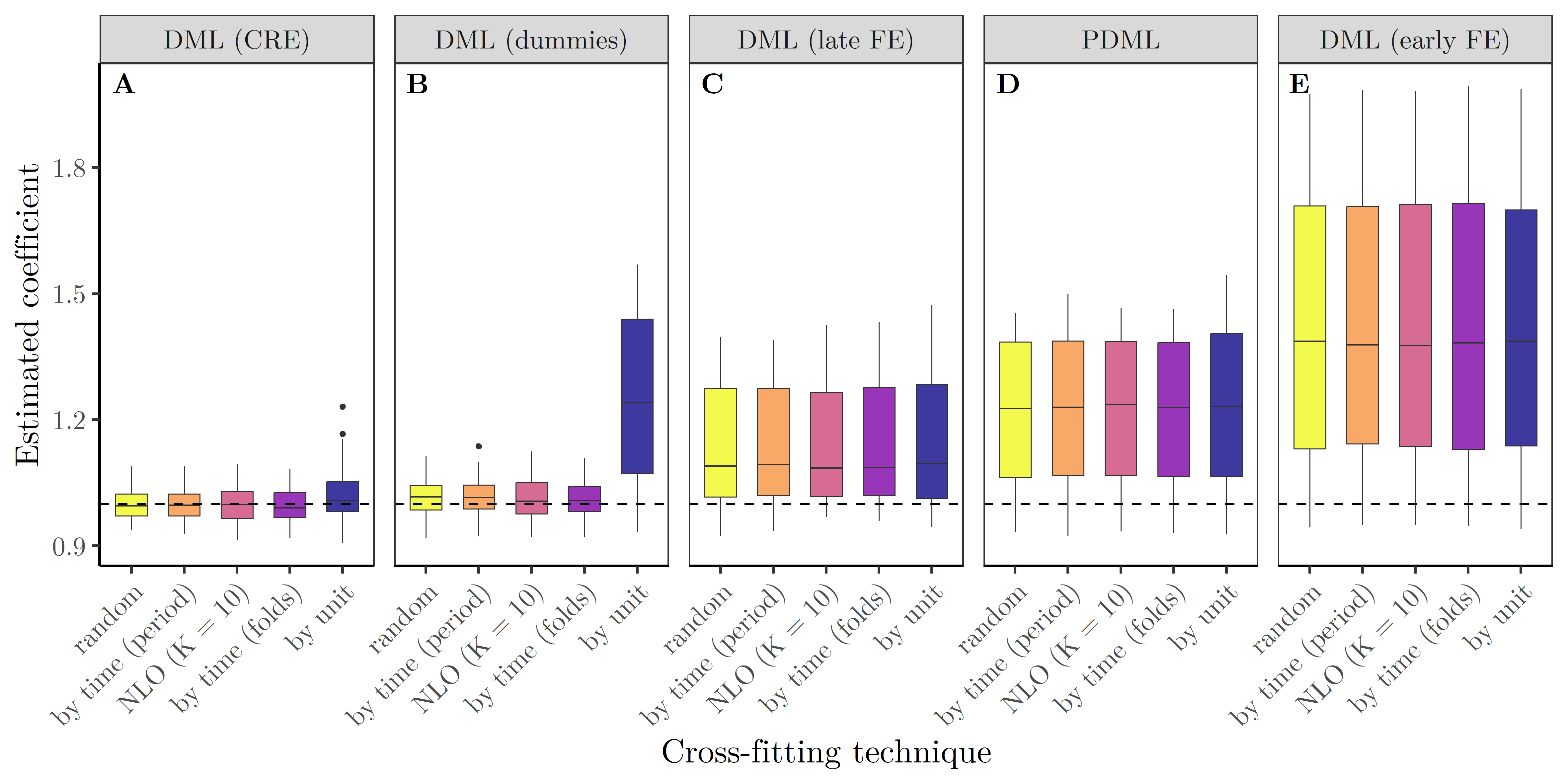}
    \caption{\label{fig:split_res}Results for utilizing different cross-fitting techniques (Table \ref{tab:split_DML_approaches}) within various DML estimators. The vertical axis depicts the estimated coefficient. The dashed line marks the true causal effect ($\beta = 1$). The boxplots show the distribution of estimated coefficients across 100 simulated datasets for each method.  Data is generated according to DGP (3), with one observed confounder, u-shaped functional forms and a large degree of autocorrelation ($\rho = 0.9$). NLO: neighbors-left-out cross-fitting.}
\end{figure}

In describing the results, we only compare the performance of different cross-fitting techniques \textit{within} the same DML estimator. In the next section, we will further describe and interpret the differences in performance between the different estimators. 
Our simulation results display a surprisingly small influence of the choice of cross-fitting technique on the estimated coefficients  (Figure \ref{fig:split_res}), except for some special cases. Within most estimation methods, the different splitting strategies lead to very similar estimated effects. One notable exception is cross-fitting when splitting by unit: as anticipated by \citet{semenova_inference_2023}, this results in substantial bias if we expect the ML methods to model the unobserved heterogeneity in the hold-out fold, while only observing the units present in the training folds. For this splitting strategy, using unit dummies within DML (Figure \ref{fig:split_res}B) leads to heavily biased effect estimates, because the dummy variables in the hold-out data belong to different units than the dummy variables used for training. To a (much) weaker degree, we make similar observations for DML with CRE (Figure \ref{fig:split_res}A) and pooled DML (Figure \ref{fig:split_res}D), though the issues become more pronounced in settings with very few units and many periods (e.g., $N = 10$ and $T = 500$, see Appendix \ref{apx:cf_resNT}). 

From these observations, we conclude that we should not split by unit when cross-fitting, at least if unobserved heterogeneity could be plausibly present in our specific application. While there is theoretical basis for ensuring independence of clustered data, especially for constructing valid confidence intervals  \citep[e.g.,][]{chiang_multiway_2022}, failing to adequately model the unobserved heterogeneity will likely lead to a more substantial bias in the estimated effects.
Besides splitting by unit, the results of the other cross-fitting techniques are hardly distinguishable. Therefore, in the absence of a strong argument for a specific cross-fitting technique, we will proceed with the rest of our simulations using the ``random" sample splitting method.

\subsection{Comparison of estimation methods}

We now compare the different estimation methods in six different baseline settings that differ in the causal structure and the functional form of the observed confounding. 
In all baseline simulations, we use $N = 500$ different units across $T = 10$ different periods. For now, we assume no autocorrelation of the error terms. 

\begin{figure}[ht]
    \centering
    \includegraphics[width=\textwidth]{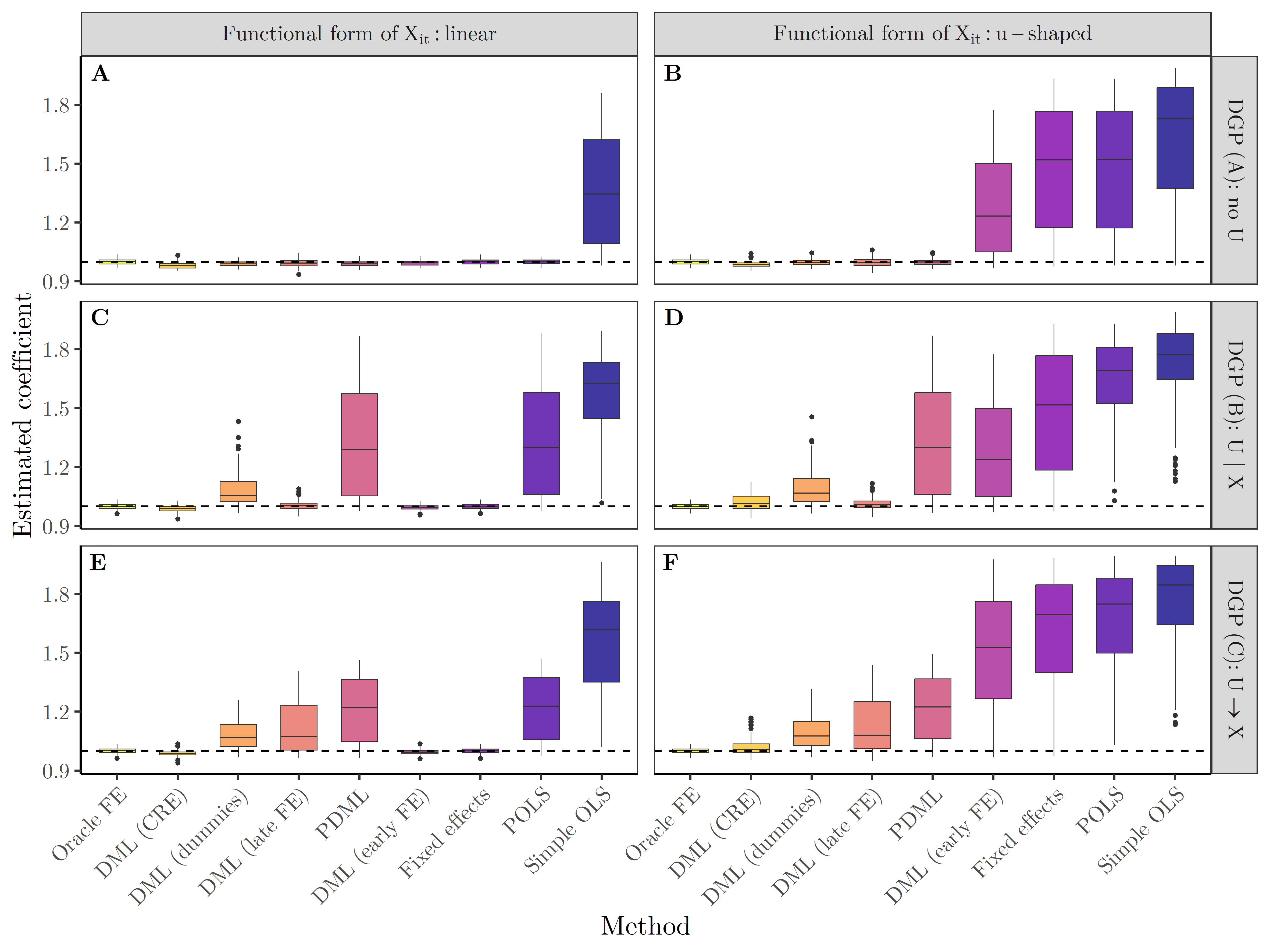}
    \caption{\label{fig:baseline}Results for our baseline simulation with $N=500$ units and $T = 10$ periods. The horizontal axis displays the different methods from Table \ref{tab:methods_impl}. The vertical axis depicts the estimated coefficient. The dashed line marks the true causal effect ($\beta = 1$). The boxplots show the distribution of estimated coefficients across 100 simulated datasets for each method. 
    The three rows contain three different DGPs: ``$no\, U$" indicates no unobserved heterogeneity, ``$U\, |\, X$" means the unobserved heterogeneity influences treatment and outcome, but not confounders, and ``$U \rightarrow X$" means the unobserved heterogeneity also influences the confounders.}
\end{figure}

The plots in the first column (Figure \ref{fig:baseline}A, C, and E) show results from simulations with a linear influence of the observed confounders $X_{it}$, whereas the results in the second column (Figure \ref{fig:baseline}B, D, and F) originate from a nonlinear, u-shaped functional form. The DGPs of the first row follow the causal structure (A), where we simulate no unobserved heterogeneity. By contrast, in the second row simulations, the unobserved heterogeneity $U_{i}$ influences only the treatment $W_{it}$ and the outcome $Y_{it}$, but not the observed confounders $X_{it}$ (causal structure (B)). The third row contains results from causal structure (C), where the unobserved heterogeneity additionally influences $X_{it}$. In terms of complexity, the simulation settings become more challenging from the top left to the bottom right panel. 

Across simulation settings, the naive simple linear regression performs worst, since it neither adjusts for potential unobserved heterogeneity nor for observed confounding. By contrast, the (infeasible) fixed effects regression with the oracle for the functional form always delivers unbiased and precise estimates.

In the simplest setting with no unobserved heterogeneity and linear confounding influence (Figure \ref{fig:baseline}A), all methods except for the naive regression perform well and give unbiased estimates. 
If we instead simulate u-shaped confounding relationships of the observed covariates $X_{it}$ (Figure \ref{fig:baseline}B), several methods become strongly biased. While this is not surprising for the linear methods (with and without fixed effects), the DML variant with early demeaning also incurs substantial bias. This is because $W_{it}$ and $Y_{it}$ are nonlinear in $X_{it}$, not in $\ddot{X_{it}} = (X_{it} - \bar{X_i})$, and the time means are not additively separable (i.e., $g_0(X_{it} - \bar{X_i}) \neq  g_0(X_{it}) - \bar{X_i}$). Hence, after demeaning, $\ddot{X_{it}}$ might not be sufficient to predict the treatment and outcome, respectively. This is especially problematic in a large $N$, small $T$ setting like our baseline: after removing the between-variation along the unit dimension ($N$), there is only little variation left along the time dimension ($T$) to model the nonlinear functional form of the observed confounding.

In the case of unobserved heterogeneity influencing treatment and outcome in addition to \textit{linear} observed confounding (Figure \ref{fig:baseline}C), we see the importance of using some sort of fixed effects estimator. Now methods such as the linear regression and pooled DML are biased, since they only adjust for the observed confounding and not for the unobserved heterogeneity. While DML with dummies performs much better, it is also not quite unbiased, because the large number of non-sparse unit dummies ($z1_i, ..., z500_i$) makes the predictions challenging, and likely leads to variable selection mistakes. All other methods deliver precise and unbiased estimates. 
Moving to nonlinear confounding within the same causal structure (Figure \ref{fig:baseline}D), the linear fixed effects estimator and DML with early demeaning become biased as well.  DML with dummies is similarly biased as in the previous setting. Of the feasible methods, only DML with correlated random effects and DML with late demeaning are close to unbiased, with the latter being slightly more precise. Since $U_i$ does not influence $X_{it}$, the unobserved heterogeneity is additively separable after modeling the observed nonlinear confounding influence, which is how the late demeaning method operates.

In the most complex causal structure considered, the unobserved heterogeneity also influences the observed confounders $X_{it}$, and thus the treatment and outcome via this second indirect path as well. In this setting with linear confounding influences, the extent of the bias for pooled OLS and pooled DML is somewhat attenuated (Figure \ref{fig:baseline}E). While these methods had no information at all about the unobserved confounding in the causal structure (B), they now can learn something about $U_i$, since it is part of $X_{it}$ via the path $U_i \rightarrow X_{it}$. For these methods, the influence of $U_i$ on $X_{it}$ is beneficial, since they can to some extent exploit that observed confounders now contain information about the unobserved heterogeneity. 
On the other hand, compared to the previous causal structure (no influence $U_i \rightarrow X_{it}$), the late demeaning method now incurs bias. 
In the $U_i \rightarrow X_{it}$ setting, this method ``overfits" treatment and outcome by modeling their relationship with $U_i$ twice. First, by predicting these variables from $X_{it}$, which now also contains information about $U_i$. Second, by demeaning the residuals after the predictions. Both residualization and demeaning try to remove the same confounding variation induced by $U_i$. Although they only remove this variation imperfectly (see, e.g., pooled OLS and pooled DML), trying to remove it twice leads to unintentionally removing exogenous variation that we need for estimating the effect of $W$ on $Y$ from the residuals, hence causing a biased estimate.

The final and most challenging setting follows the same causal structure, but uses a nonlinear influence of the observed confounders (Figure \ref{fig:baseline}F). As in Panel E, the late demeaning method again cannot deliver unbiased estimates. The only feasible method without substantial bias is DML with correlated random effects, followed by DML with dummies, which incurs a similar bias as in the previous three settings. Compared to Panel D, pooled DML is less biased, since it (partially) learns about $U_i$ through $X_{it}$.

Across settings, only DML with correlated random effects consistently delivers estimates that are close to the true effect. If the influence of the observed confounders is linear, standard fixed effects estimation is appropriate and sufficient, but it fails if that influence is different from the one specified in the fixed effects model (e.g., nonlinear instead of linear). If we can rule out an influence of the unobserved heterogeneity on the observed confounders, DML with late demeaning gives very accurate estimates. However, since we rarely can rule out that influence with certainty in practice (or have reason to believe that it exists, like in our example), our simulations suggest that DML with correlated random effects is most robust to any of the considered baseline settings.

\subsection{Simulation extensions}

From the initial baseline, we vary different characteristics of the DGP to explore how the suggested methods perform in a variety of settings.

\subsubsection{Changing the panel dimensions  \texorpdfstring{$N$}{N}/ \texorpdfstring{$T$}{T}}

First, we vary the relation of the number of units ($N$) to the number of time periods ($T$), while keeping the overall number of observations constant. In addition to the baseline setting [$N = 500$, $T = 10$], we also consider the combinations [$N = 100$, $T = 50$], [$N = 50$, $T = 100$], and [$N = 10$, $T = 500$]. Since we currently only simulate unobserved heterogeneity on the unit dimension and not on the time dimension, we expect the smaller numbers of units (and thus decreased dimensionality) to benefit the DML with dummies method most of all. 

\begin{figure}[ht]
    \centering
    \includegraphics[width=\textwidth]{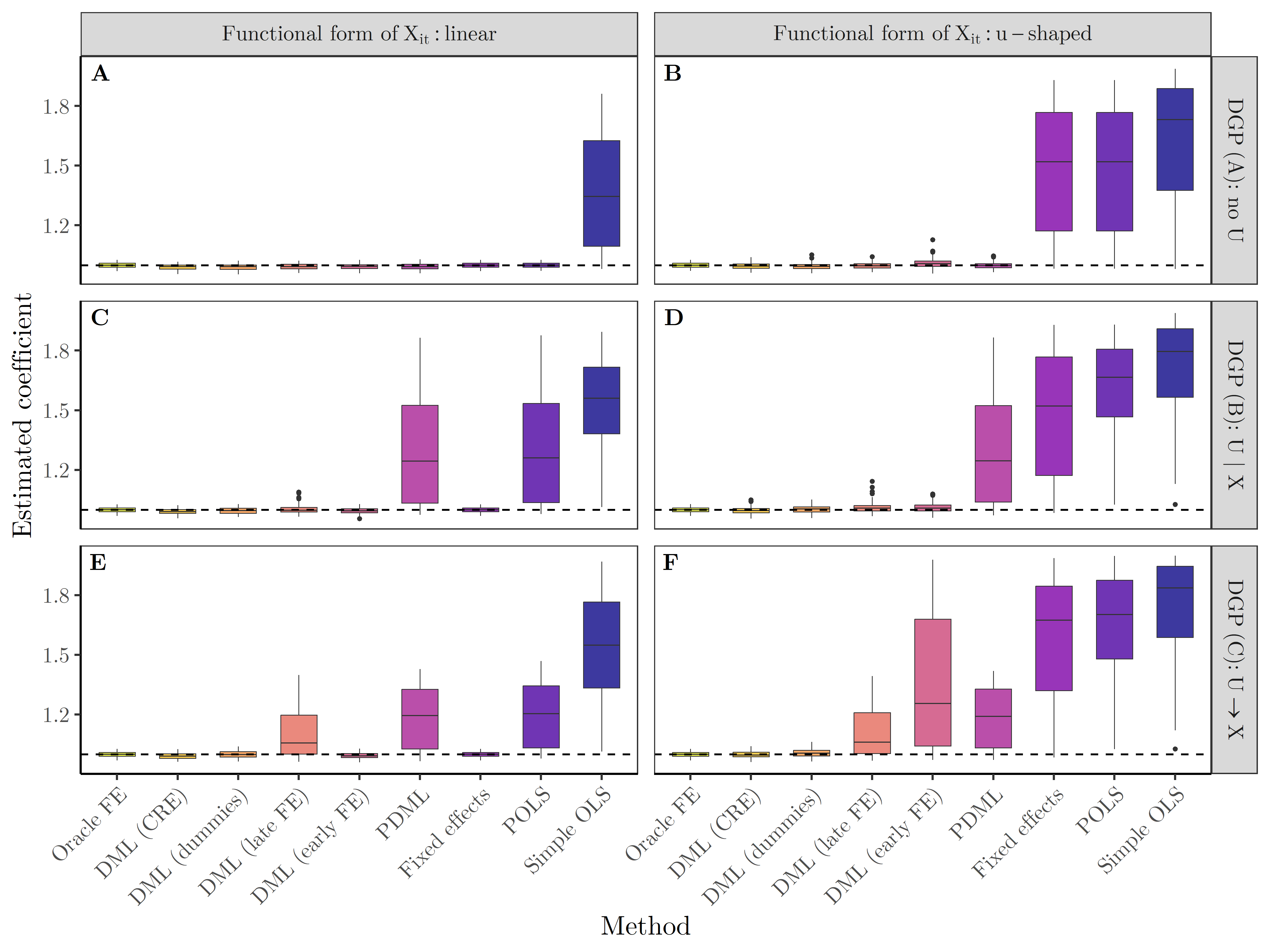}
    \caption{\label{fig:n10t500}Results for the setting with $N=10$ units and $T = 500$ periods. 
    The dashed line marks the true causal effect ($\beta = 1$). The boxplots show the distribution of estimated coefficients across 100 simulated datasets for each method. 
    The three rows contain three different DGPs: ``$no\, U$" indicates no unobserved heterogeneity, ``$U\, |\, X$" means the unobserved heterogeneity influences treatment and outcome, but not confounders, and ``$U \rightarrow X$" means the unobserved heterogeneity also influences the confounders.}
\end{figure}

We report results for the setting with $N = 10$ units and $T = 500$ periods (Figure \ref{fig:n10t500}). In comparison to our baseline (Figure \ref{fig:baseline}), the estimates of two particular methods change substantially. First, using DML with dummy variables now results in virtually unbiased estimates. The ML algorithm only has to handle 10 dummy variables (one for each unit) instead of the 500 dummies in the baseline scenario. This easier task does not result in variable selection mistakes, and thus the method estimates the effect almost as precisely as DML with correlated random effects. 
Second, DML with early demeaning now also delivers precise estimates in the less complex nonlinear settings (Figure \ref{fig:n10t500}B and D), but still not for the complex setting in Panel F.
As mentioned above, the early demeaning removes unit-specific between-variation, which the ML method subsequently cannot use for the prediction task. However, in the absence of  $U_i \rightarrow X_{it}$, this is only consequential if the between-variation is crucial for modeling the confounding relationships. In the current small $N$, large $T$ setting, the within-variation still consists of 500 observations per unit, which is sufficient for modeling the nonlinear functional forms. 
In the more complex setting (Panel F), the early demeaning still cannot fully remove the unobserved heterogeneity, since the time means are not additively separable due to the nonlinear function: $g_0(X_{it} - \bar{X_i}) \neq  g_0(X_{it}) - \bar{X_i}$. 
The performance of the remaining methods is very similar to the baseline setting.

The settings with [$N = 100$, $T = 50$] and  [$N = 50$, $T = 100$] lead to results between the baseline and the [$N = 10$, $T = 500$] setting, such that  DML with dummies and DML with early demeaning decline in accuracy as $N$ increases and $T$ decreases (see Appendix \ref{apx:furth_resNT}).

\subsubsection{Increasing the number of observed confounders}

In the second extension, we test how the different methods scale for larger numbers of observed confounders $\boldsymbol{X_{it}}$. We focus on causal structure (C) with nonlinear confounding and generate $X_{jit}$ for $j = 1, ..., J$ confounders according to Equation \ref{eq:sim_nconfx}, where we draw $\epsilon_{jit}$  from a multivariate normal distribution with a mean of zero and a randomly generated covariance matrix (i.e., $\epsilon_{jit} \sim N(0, \Sigma)$, $\Sigma = A'A$, with $A \sim N(0, 1)$).
\begin{equation} \label{eq:sim_nconfx}
X_{jit} = \alpha_{0j} + \delta U_i + \epsilon_{jit}
\end{equation}
Also, we now draw separate confounding coefficients $\gamma_j$ for each confounder and divide each by the overall number of confounders $J$ (Equations \ref{eq:sim_nconfw} and \ref{eq:sim_nconfy}). In doing so, we ensure that on average, the overall strength of the confounding influence is similar to the baseline scenario, and that we are only varying the number of confounders. 
\begin{gather}  
W_{it} = \alpha_1 + \sum_{j = 1}^{J}  \frac{\gamma_j}{J} g_0(X_{jit}) + \delta U_i + \eta_{it} \label{eq:sim_nconfw} \\
Y_{it} = \alpha_2 + \beta W_{it} + \sum_{j = 1}^{J} \frac{\gamma_j}{J}  m_0(X_{jit}) + \delta U_i + \mu_{it} \label{eq:sim_nconfy}
\end{gather}
For this setting and the next, we introduce two new methods as baselines to facilitate the interpretation of the results. First, the method ``FE only" does not adjust for the observed confounding ($\boldsymbol{X_{it}}$) at all, but only aims to account for unobserved heterogeneity by using fixed effects. This method allows us to determine how important directly adjusting for the unobserved heterogeneity is. 
Second, the (infeasible) method ``Oracle w/o FE" adjusts for the observed confounding by knowing the ``true" functional form, but does not account for the unobserved heterogeneity, thereby demonstrating the significance of adjusting for the $\boldsymbol{X_{it}}$ only. Finally, the additional benefit of also adjusting for the unobserved heterogeneity becomes visible in the previously described ``Oracle FE" method. 

\begin{figure}[ht]
    \centering
    \includegraphics[width=\textwidth]{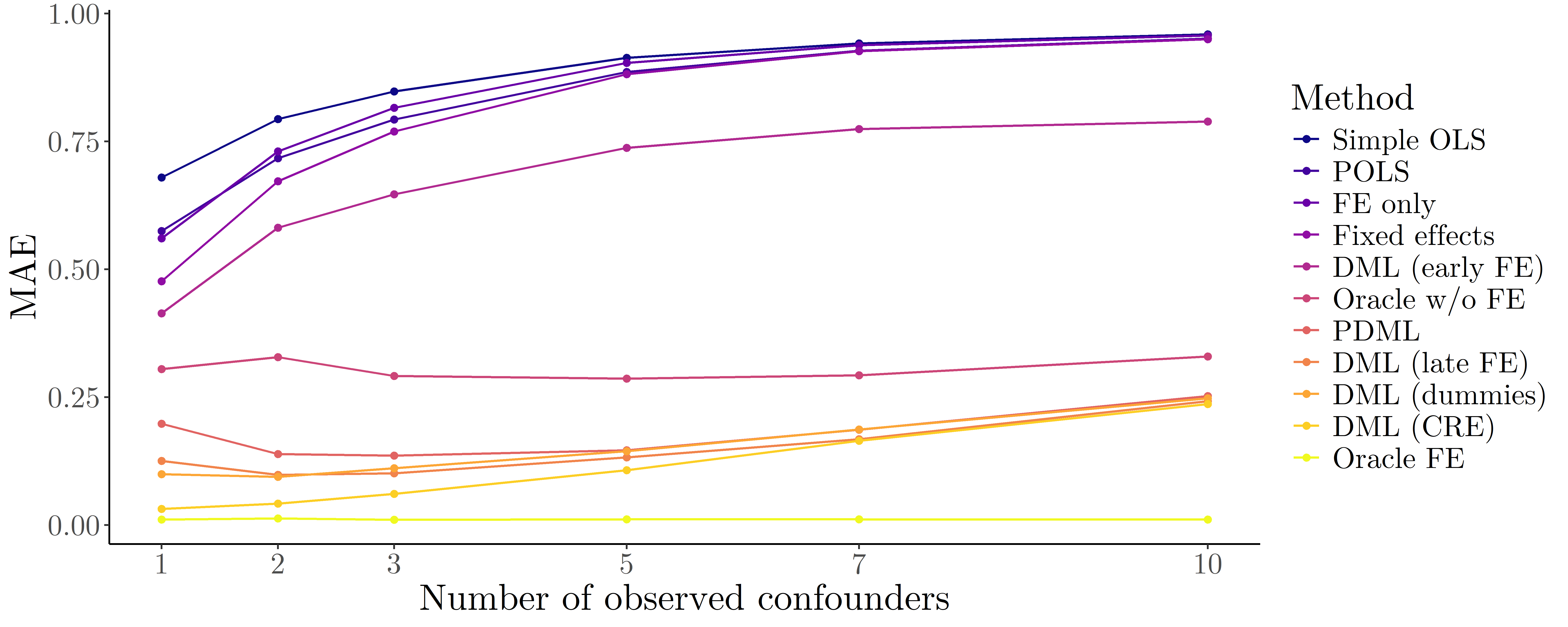}
    \caption{\label{fig:nconf}Mean absolute error in the estimated coefficient across 100 simulations by number of observed confounders. The simulated influence of the observed confounders is u-shaped, the causal structure is (C), i.e., $U_i \rightarrow X_{it}$. $N = 500$, $T = 10$.}
\end{figure}

We display the resulting mean absolute error (MAE) in the causal coefficient for each method for between 1 and 10 observed confounders (Figure \ref{fig:nconf}). We first observe that the (infeasible) oracle methods are barely affected by changing the number of confounders: the full oracle (``Oracle FE") perfectly adjusts for both observed and unobserved confounding and remains unbiased for all numbers of confounders, while only adjusting perfectly for observed confounding (``Oracle w/o FE") leads to an almost constant bias around 30\%.

Contrary to the oracle methods, virtually all other approaches increase in MAE as the number of confounders increases. Also, the gap between methods explicitly accounting for unobserved heterogeneity and their counterpart that does not becomes smaller or vanishes for larger numbers of confounders. For example, ``FE only" incurs a similar degree of bias as ``Simple OLS" after 5 confounders. 
Comparing the DML methods (excluding early demeaning) to the functional form oracle, we observe that they all are substantially more accurate
. This indicates that these methods do not only adjust for the observed confounding well (as ``Oracle w/o FE" does), but also capture parts of the unobserved heterogeneity. 
Interestingly, the pooled DML approach outperforms the functional form oracle as well, even though it has no explicit way of adjusting for the unobserved heterogeneity. This is because PDML can \textit{indirectly} adjust for part of the unobserved heterogeneity, as $\boldsymbol{X_{it}}$ contains variation caused by $U_i$ ($U_i \rightarrow \boldsymbol{X_{it}}$), which facilitates the prediction of treatment and outcome.
However, as the number of confounders and thus the dimensionality increases at a constant sample size, all DML methods after some point struggle to still adjust effectively. The pooled DML and ``late" fixed effects approaches improve from one to two confounders, before increasing in bias similar to the others methods for more confounders. 
These approaches can benefit from multiple  $\boldsymbol{X_{it}}$: as the dimensionality of $\boldsymbol{X_{it}}$ increases, the direct influence of the unobserved $U_i$ becomes less important, while the indirect influence is blocked by adjusting for the $\boldsymbol{X_{it}}$. This benefit of having multiple observed confounders in the $U_i \rightarrow \boldsymbol{X_{it}}$ case becomes even more evident when the prediction tasks are simpler, e.g., for linear confounding or larger sample sizes (see Figures \ref{fig:nconf_linear} and \ref{fig:nconf500_100}, respectively, in the Appendix).
At the same time, adjustment for further $\boldsymbol{X_{it}}$ becomes more challenging in our baseline setting, hence the negative effects of additional confounders quickly dominate and the bias increases.

Using fixed effects late in DML and using dummies behave relatively similarly, with the dummy approach scaling slightly worse with the number of confounders. 
While DML with the CRE approach is almost unbiased for one confounder, its bias increases with the number of confounders more quickly than that of the other methods. From seven confounders on, it incurs  bias similar to the late fixed effects method; for ten confounders, it is similar to the pooled DML. This is likely because the dimensionality in the CRE approach increases by a factor of $2 * J$, as for each observed confounder we have to adjust both for the time-varying variable and its time mean. By contrast, the other methods only have to handle $J$ variables, making the adjustment process less complex. 
While this appears to be a downside of the DML with CRE method, in the next set of simulations we investigate whether the method can handle larger numbers of confounders, provided a sufficiently large sample size.

\subsubsection{Varying the sample size}

We now assess how the sample size influences the estimates of the different methods. Is the bias in the estimates only a consequence of finite samples or is there some systematic issue with the estimators that prevents them from getting closer to the true effect?  
The results show that while most feasible methods are not guaranteed to substantially improve with sample size, DML with CRE can be unbiased as long as the number of observations is large relative to the number and strength of the observed confounders. 

\begin{figure}[ht]
    \centering
    \includegraphics[width=\textwidth]{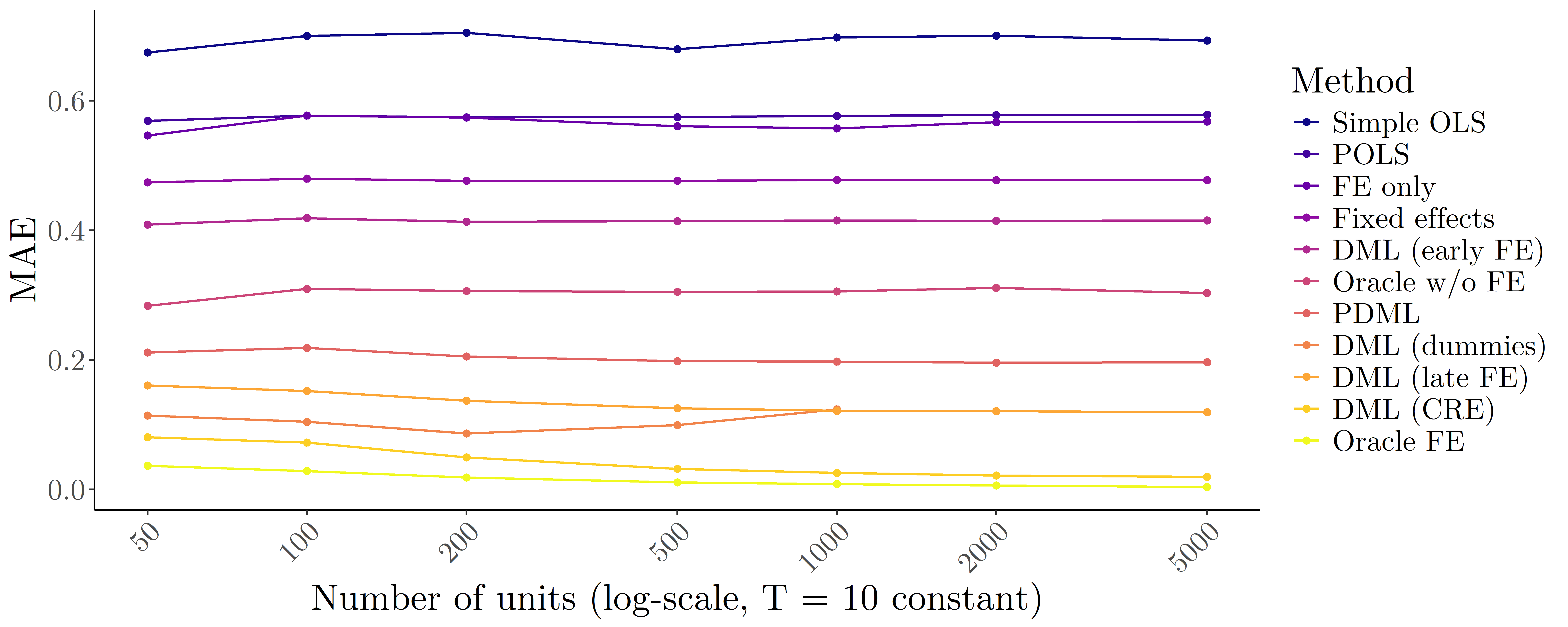}
    \caption{\label{fig:ss1N_nconf}Mean absolute error in the estimated coefficient across 100 simulations for \textbf{1} observed confounder by the number of units. The number of periods is fixed at $T = 10$. The simulated influence of the observed confounder is u-shaped, the causal structure is (C), i.e., $U_i \rightarrow X_{it}$. We computed DML (dummies) only for up to $N = 1000$, as it becomes computationally too costly for larger values.}
\end{figure}

First, we vary the number of units $N$ in the baseline setting with causal structure (C) and one observed confounder with a u-shaped functional form (Figure \ref{fig:ss1N_nconf}). DML with CRE is the only feasible method that is close to the oracle method, and it gets more precise as the number of observations increases. None of the other DML-based methods substantially improve in larger samples, since they systematically lack the ability to model the unobserved heterogeneity. 

\begin{figure}[ht]
    \centering
    \includegraphics[width=\textwidth]{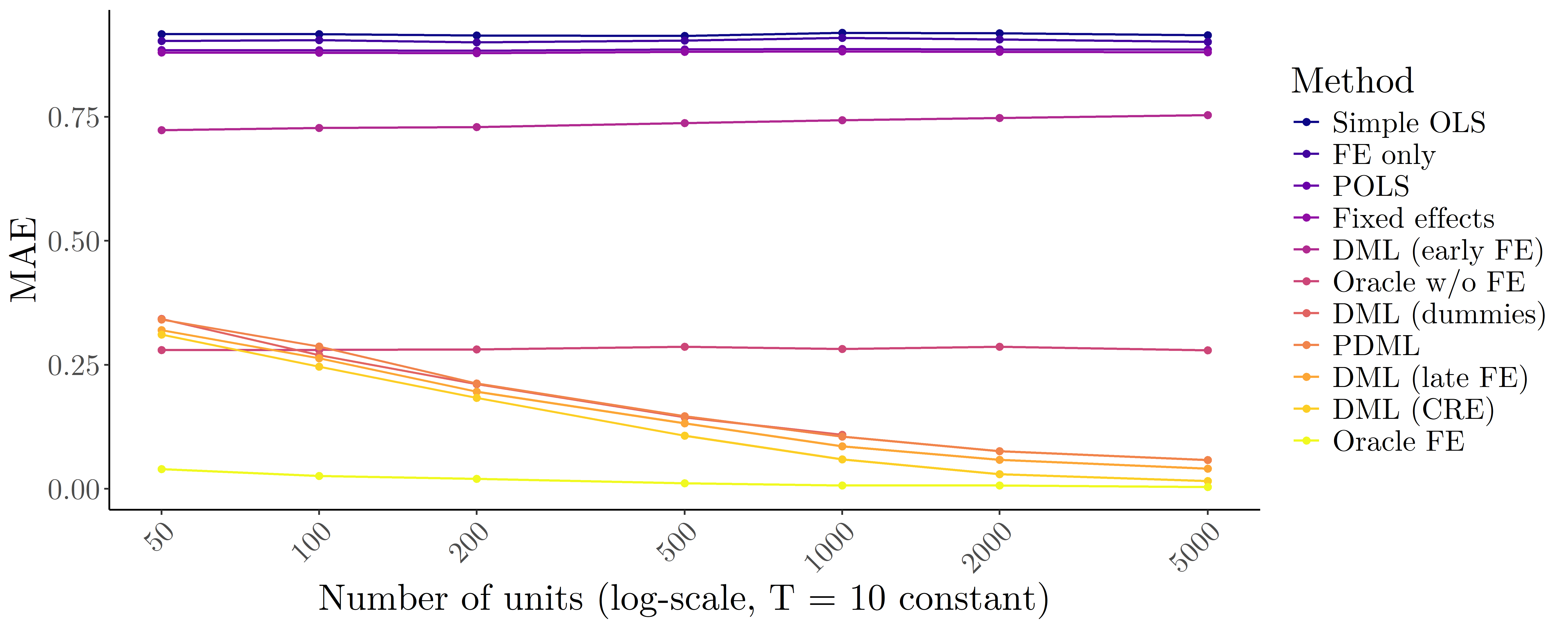}
    \caption{\label{fig:ss5N_nconf}Mean absolute error in the estimated coefficient across 100 simulations for \textbf{5} observed confounders by the number of units. The number of periods is fixed at $T = 10$. The simulated influence of the observed confounders is u-shaped, the causal structure is (C), i.e., $U_i \rightarrow X_{it}$. We computed DML (dummies) only for up to $N = 1000$, as it becomes computationally too costly for larger values.}
\end{figure}

This behavior changes when we look at a setting with more observed confounders (Figure \ref{fig:ss5N_nconf}; 5 confounders). Now the other DML-based methods (except for the early demeaning) substantially improve as the sample size increases, since the additional observed confounders contain more information about the unobserved heterogeneity. However, DML with CRE remains dominant at every sample size, even extends its advantage for very large samples, and converges to the oracle method for $N = 5000$.

We observe similar results for settings where we vary the number of periods $T$ while keeping the number of units constant at $N = 500$ (see Figures \ref{fig:ss1T_nconf} and \ref{fig:ss5T_nconf} in the Appendix). As the only noticeable difference, DML with early demeaning there also improves as the number of periods (i.e., the amount of within-variation) increases, although it is still heavily biased even at $T = 400$.

\subsubsection{Two-way fixed effects}

Here, we augment the unit-specific unobserved heterogeneity by also including time-specific unobserved heterogeneity. Hence, the DGP in Equations \ref{eq:sim_conftwfe}-\ref{eq:sim_outtwfe} now also contains the unobserved variable $U_t$, which varies only over time. In the estimation process, methods using fixed effects now employ \textit{both} unit and time fixed effects, while the correlated random effects approach includes \textit{both} unit and time means for treatment and covariates, in addition to the original covariates.  
\begin{gather} 
X_{it} = \alpha_0 + \delta U_i + \delta U_t + \epsilon_{it} \label{eq:sim_conftwfe} \\
W_{it} = \alpha_1 +  \gamma g_0(X_{it}) +  \delta U_i + \delta U_t + \eta_{it} \label{eq:sim_treattwfe} \\
Y_{it} = \alpha_2 + \beta W_{it} + \gamma m_0(X_{it}) + \delta U_i + \delta U_t + \mu_{it}  \label{eq:sim_outtwfe} 
\end{gather}
\begin{figure}[ht]
    \centering
    \includegraphics[width=\textwidth]{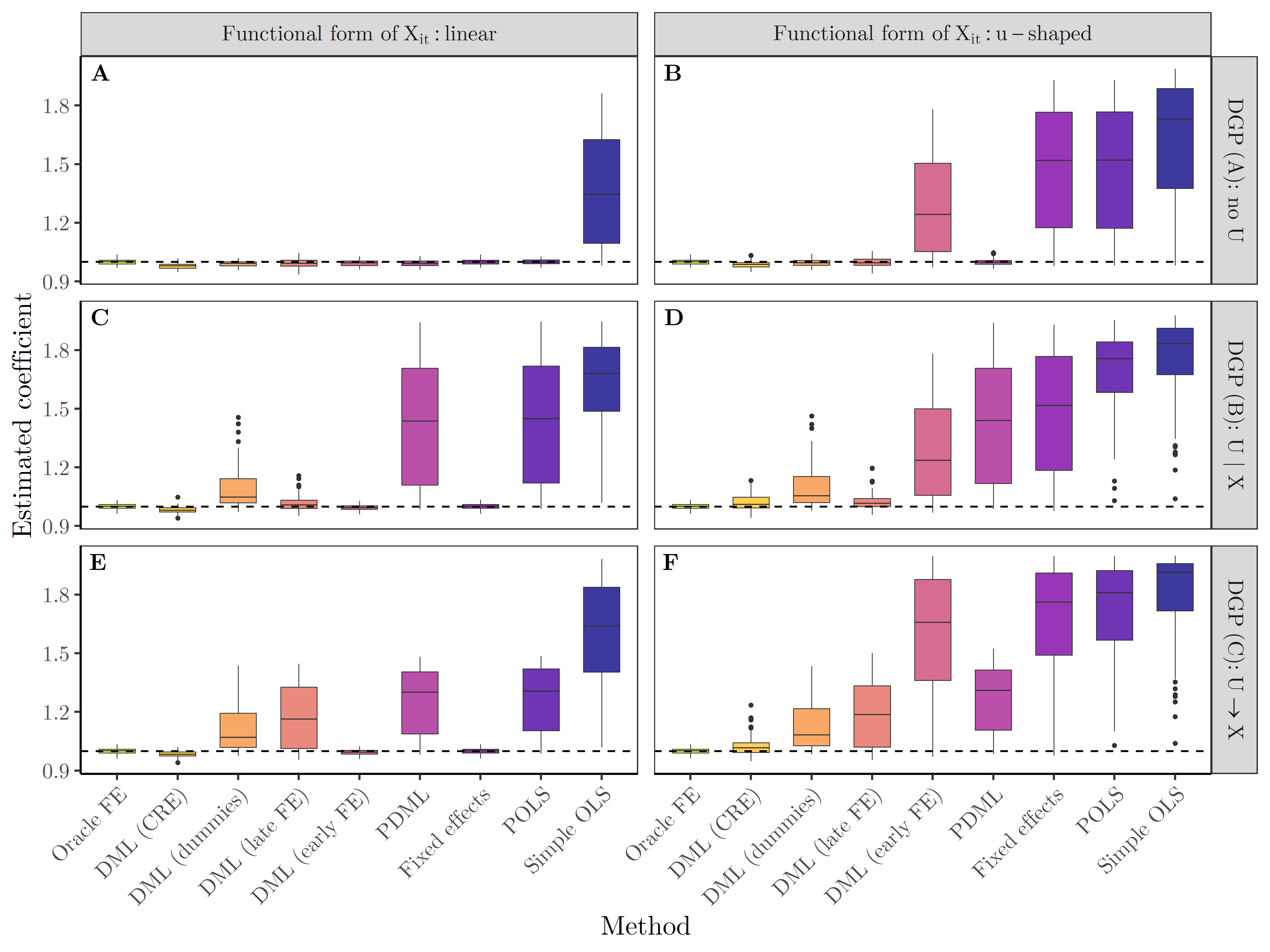}
    \caption{\label{fig:twfe}Results for DGPs with unobserved heterogeneity in both the unit and the time dimension with $N=500$ units and $T = 10$ periods. The dashed line marks the true causal effect ($\beta = 1$). The boxplots show the distribution of estimated coefficients across 100 simulated datasets for each method. 
    The three rows contain three different DGPs: ``$no\, U$" indicates no unobserved heterogeneity, ``$U\, |\, X$" means the unobserved heterogeneity influences treatment and outcome, but not confounders, and ``$U \rightarrow X$" means the unobserved heterogeneity also influences the confounders.
    }
\end{figure}
Compared to the baseline setting, no method performs substantially different under two-way fixed effects (Figure \ref{fig:twfe}). All methods that cannot fully account for the unobserved heterogeneity in the baseline now perform slightly worse, since there is additional time-varying heterogeneity they can also not account for. DML with correlated random effects seem virtually unaffected by the added dimension.

\subsubsection{Autocorrelation}

One relevant difference between cross-sectional and panel data is the potential for autocorrelation (or serial correlation) in the latter, i.e., the error terms of the outcome model could be correlated across time (\citealp{wooldridge_introductory_2012}, Chapter 10). Serial correlation does not prevent consistency and unbiasedness in OLS under strict exogeneity (\citealp[]{wooldridge_introductory_2012}, Chapter 12), but invalidates the usual OLS standard errors (even though cluster robust standard errors are available for traditional methods). While standard errors are not the focus of our analysis, we want to explore whether serial correlation can also become problematic for the estimated coefficients in DML, where it violates the i.i.d.\ assumption in the cross-fitting procedure. We already explored the impact of different cross-fitting procedures on the estimates in the presence of autocorrelation in Section \ref{comp_cf}. Now, we investigate how different degrees of autocorrelation affect the estimates in our baseline setting with causal structure (C). 

To test the methods' sensitivity to autocorrelation, we change the DGP of the outcome model (Equation \ref{eq:sim_out}) to include serially correlated errors: Now, we simulate $\mu_{it}$ according to an AR(1) model, i.e., $\mu_{it} = \rho\mu_{it-1} + e_{it}$, with the ar-coefficient $\rho$ equal to 0, 0.5, or 0.9, implying no, medium, and substantial autocorrelation, respectively. We allow for a longer time series than in our baseline by going back to the setting with $N = 100$ observations and $T = 50$ periods. 

\begin{figure}[ht]
    \centering
    \includegraphics[width=\textwidth]{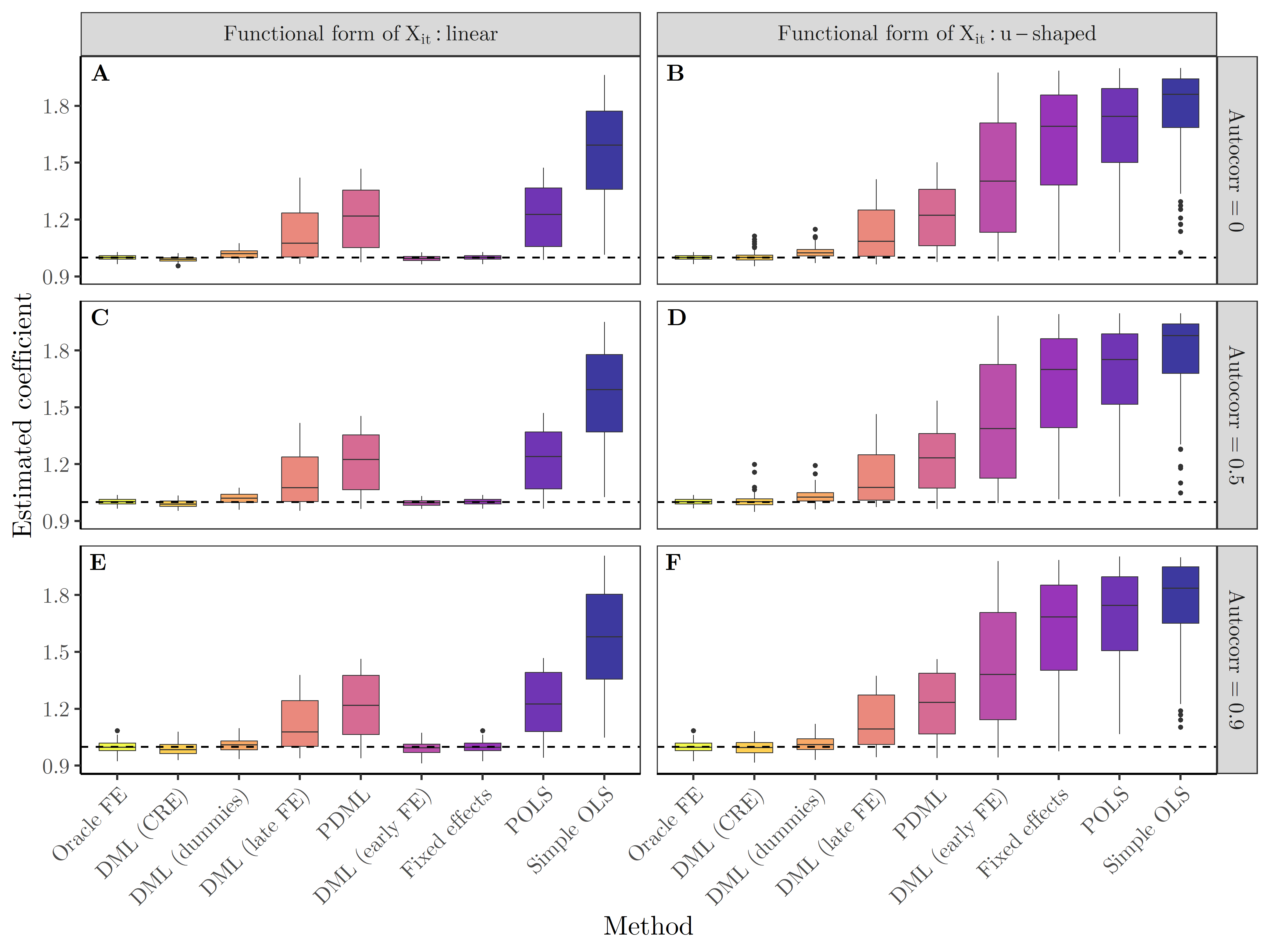}
    \caption{\label{fig:autocorr}Results for different degrees of autocorrelation with $N=100$ units and $T = 50$ periods. Panel A, B: no autocorrelation; Panel C, D: medium degree of autocorrelation; Panel E, F: strong autocorrelation.     The dashed line marks the true causal effect ($\beta = 1$). The boxplots show the distribution of estimated coefficients across 100 simulated datasets for each method. The simulated causal structure is (C), i.e., $U_i \rightarrow X_{it}$. }
\end{figure}

We find that even larger degrees of autocorrelation do not substantially alter our conclusions about the estimated coefficients (Figure \ref{fig:autocorr}). Only the distribution of the estimates of the well-performing method becomes slightly wider, while still being centered around very similar values. We conclude that in DML, the degree of autocorrelation is not critical for the accuracy of the estimated coefficients, and will probably only complicate the construction of valid standard errors.

\section{Discussion} \label{sec:disc}

In this article, we explored how we can adapt the double/debiased machine learning framework to deal with unobserved heterogeneity in panel data settings. For this purpose, we considered several intuitive methods to account for unit-specific heterogeneity within DML. If DML can thereby adjust for time-constant unobserved confounding, similarly to the traditional fixed effects estimator, it allows us to relax \textit{two} assumptions in the estimation of causal effects: (1) the assumption that we have chosen the correct functional forms in a parametric model, since a flexible ML method can in principle learn these within DML, and (2) the assumption that there is no unobserved confounding, since accounting for unobserved heterogeneity allows us to rely on the weaker assumption of no \textit{time-varying} unobserved confounding. 
However, adaptations of DML to panel data settings are not straightforward due to two issues. First, the dependence of observations within the same cluster and/or across time violates assumptions underlying the cross-fitting procedure of DML. Secondly, nonlinear confounding relationships complicate the elimination of unobserved heterogeneity.

While our results show that the cross-fitting procedure is not crucial for the accuracy of the estimated effects, we also find that many of the intuitive methods fail to recover the true effect in simulated data with nonlinear influences of the observed confounders. 
Although most DML-based methods are superior to estimating a misspecified (i.e., linear) fixed effects model in such settings, most of them cannot fully remove the confounding bias. 
We demonstrate that the influence of the unobserved unit-specific confounding on the observed time-varying confounders ($U_i \rightarrow X_{it}$) plays a critical role in the viability of some methods: due to the nonlinear influence of the observed confounders, the unobserved heterogeneity is no longer additively separable and most approaches cannot easily eliminate it. The only DML estimator delivering good estimates across settings is the one using the Mundlak-type correlated random effects approach within DML: by using the time-means of the treatment and the covariates as additional predictors within DML, this approach can explicitly model the unobserved heterogeneity, even in cases with nonlinear observed confounding. One caveat of this estimator is the need for a sample size that is large relative to the number of observed confounders, which however is not unreasonable in many applications. 

We next discuss whether it is reasonable to stress the importance of the influence of the unobserved heterogeneity on the observed confounding ($U_i \rightarrow X_{it}$) like we do in this article. Is this path likely to be present in typical applications? We consider two classical causal questions as examples: the evaluation of job programs, and the estimation of the price elasticity of demand for consumer goods. First, one standard problem in econometric textbooks is estimating the effect of a job program ($W_{it}$) on future earnings of the participants ($Y_{it}$) \citep[e.g.,][Chapter 10]{wooldridge_econometric_2010}. Important unobserved, but (relatively) time-constant confounders ($U_i$) could be the ability or motivation of individuals, which could influence both whether they participate in the training and their future earnings. Observed confounders typically consist of individual characteristics like age, sex, years of schooling, marital status, number of hours worked before the training, etc. While some of these are certainly not influenced by the unobserved heterogeneity (age, sex) and would drop out of a fixed effects regression anyway, others can vary over time (schooling, marital status, hours worked) and are plausibly influenced by the ability and/or motivation ($U_i \rightarrow X_{it}$). 
Secondly, we return to our example from the introduction. Because pricing decisions are very consequential for many products, retailing firms want to know the effect of price ($W_{it}$) on demand ($Y_{it}$) \citep[e.g.,][]{bijmolt_new_2005}. However, if the available panel data consists of different stores over time, there could be unobserved heterogeneity ($U_i$), e.g., due to management quality, the attractiveness of the store location, or demographic characteristics of residents nearby \citep[e.g.,][]{papies_addressing_2017, wooldridge_econometric_2010}. At the same time, observed factors like advertising, sales promotions, or prices of competitor products could function as time-varying confounding variables ($X_{it}$), and are likely also influenced by the unobserved time-constant factors ($U_i \rightarrow X_{it}$). In sum, these two examples illustrate that the influence of $U_i$ on $X_{it}$ is arguably common in practice. As a consequence, researchers should ensure that the estimators they use are capable of dealing with this setting, like we show for DML with CRE. 

From our results, we derive the following recommendations: (1) In applications where researchers currently use the traditional fixed effects regression, we recommend additionally using DML with correlated random effects and comparing the estimated coefficients. This robustness check can indicate potential for functional form misspecification if the estimated effects are very different. (2) When comparing the results, we recommend making the confidence in the DML estimate dependent on the sample size. We encourage a higher confidence in the accuracy of the estimated effect if the sample size is large and the number of observed confounders is small. (3) We recommend \textit{not} splitting by unit within cross-fitting, at least if the goal is to model the unobserved heterogeneity within the ML prediction steps. (4) We discourage interpreting standard errors or confidence intervals stemming from our considered DML estimators. There is still a lack of clarity about how to obtain valid variance estimators in these settings with both cluster and time dependence, especially in combination with the cross-fitting procedure. 

Our final recommendation already ties in with the limitations of our study. We focus only on whether the suggested estimators are capable of retrieving the true causal effect in a variety of simulated settings. We do not provide any asymptotic properties of these estimators and are thus not able to construct valid confidence intervals. Also, our considered estimators are motivated by intuitively appealing adaptations of existing panel data methods and not built on Neyman-orthogonal scores like the ones in \citet{chernozhukov_doubledebiased_2018}. By contrast, \citet{clarke_double_2023} derive a Neyman-orthogonal score for a similar setting, but also show inference to be problematic in a number of situations. Further exploration of the possibilities and limitations of statistical inference in these settings will be crucial for a more widespread adoption of these methods in practical applications.  
Finally, we are assuming the absence of dynamic effects and effect heterogeneity, and only consider additive fixed effects. Future research could study the consequences of violating these assumptions and propose alternative DML adaptations for these situations.

\section*{Acknowledgements}
Funded by the Deutsche Forschungsgemeinschaft (DFG, German Research Foundation) under Germany’s Excellence Strategy – EXC number 2064/1 – Project number 390727645. 
The authors acknowledge support by the state of Baden-Württemberg through bwHPC.

\clearpage
\bibliography{dml_panel}
\bibliographystyle{apalike}
\clearpage

\appendix
\section{Appendix} \label{appendix}

\subsection{Cross-fitting techniques illustrations and results}

\subsubsection{Illustration of splitting procedures} \label{apx:split_illus}

\begin{figure}[ht!]
    \centering
    \begin{tikzpicture}
    \node (B) at (-2,0)  {};
    \node (B) at (-1,0)  {$\boldsymbol{\mathcal{M}_1}$};
    \node (B) at (0,0)  {$\mathcal{M}_2$};
    \node (B) at (1,0)  {$\mathcal{M}_3$};
    \node (B) at (2,0)  {$\mathcal{M}_4$};
    \node (B) at (3,0)  {$\mathcal{M}_5$};
    \node (B) at (4,0)  {$\mathcal{M}_6$};
    \node (B) at (5,0)  {$\mathcal{M}_7$};
    \node (B) at (6,0)  {$\mathcal{M}_8$};
    \node (B) at (7,0)  {$\mathcal{M}_9$};
    \node (B) at (8,0)  {$\mathcal{M}_{10}$};
    \node (box) at (-1.5,0) [draw = ut-dark, minimum width = 1cm, minimum height = .5cm, anchor = west] {};
    \node (box2) at (4.1,0) [draw = ut-gold, minimum width = 8.9cm, minimum height = .5cm] {};
\end{tikzpicture}

\begin{tikzpicture}
    \node (B) at (-2,0)  {};
    \node (B) at (-1,0)  {$\mathcal{M}_1$};
    \node (B) at (0,0)  {$\boldsymbol{\mathcal{M}_2}$};
    \node (B) at (1,0)  {$\mathcal{M}_3$};
    \node (B) at (2,0)  {$\mathcal{M}_4$};
    \node (B) at (3,0)  {$\mathcal{M}_5$};
    \node (B) at (4,0)  {$\mathcal{M}_6$};
    \node (B) at (5,0)  {$\mathcal{M}_7$};
    \node (B) at (6,0)  {$\mathcal{M}_8$};
    \node (B) at (7,0)  {$\mathcal{M}_9$};
    \node (B) at (8,0)  {$\mathcal{M}_{10}$};
    \node (box0) at (-1.5,0) [draw = ut-gold, minimum width = .9cm, minimum height = .5cm, anchor = west] {};
    \node (box) at (0,0) [draw = ut-dark, minimum width = 1cm, minimum height = .5cm] {};
    \node (box2) at (4.6,0) [draw = ut-gold, minimum width = 7.9cm, minimum height = .5cm] {};
\end{tikzpicture}

\begin{tikzpicture}
    \node (B) at (-2,0)  {};
    \node (B) at (-1,0)  {$\mathcal{M}_1$};
    \node (B) at (0,0)  {$\mathcal{M}_2$};
    \node (B) at (1,0)  {$\boldsymbol{\mathcal{M}_3}$};
    \node (B) at (2,0)  {$\mathcal{M}_4$};
    \node (B) at (3,0)  {$\mathcal{M}_5$};
    \node (B) at (4,0)  {$\mathcal{M}_6$};
    \node (B) at (5,0)  {$\mathcal{M}_7$};
    \node (B) at (6,0)  {$\mathcal{M}_8$};
    \node (B) at (7,0)  {$\mathcal{M}_9$};
    \node (B) at (8,0)  {$\mathcal{M}_{10}$};
    \node (box0) at (-1.5,0) [draw = ut-gold, minimum width = 1.9cm, minimum height = .5cm, anchor = west] {};
    \node (box) at (1,0) [draw = ut-dark, minimum width = 1cm, minimum height = .5cm] {};
    \node (box2) at (5.1,0) [draw = ut-gold, minimum width = 6.9cm, minimum height = .5cm] {};
\end{tikzpicture}

\begin{tikzpicture}
    \node (B) at (-2,0)  {};
    \node (B) at (-1,0)  {$\mathcal{M}_1$};
    \node (B) at (0,0)  {$\mathcal{M}_2$};
    \node (B) at (1,0)  {$\mathcal{M}_3$};
    \node (B) at (2,0)  {$\boldsymbol{\mathcal{M}_4}$};
    \node (B) at (3,0)  {$\mathcal{M}_5$};
    \node (B) at (4,0)  {$\mathcal{M}_6$};
    \node (B) at (5,0)  {$\mathcal{M}_7$};
    \node (B) at (6,0)  {$\mathcal{M}_8$};
    \node (B) at (7,0)  {$\mathcal{M}_9$};
    \node (B) at (8,0)  {$\mathcal{M}_{10}$};
    \node (box0) at (-1.5,0) [draw = ut-gold, minimum width = 2.9cm, minimum height = .5cm, anchor = west] {};
    \node (box) at (2,0) [draw = ut-dark, minimum width = 1cm, minimum height = .5cm] {};
    \node (box2) at (5.6,0) [draw = ut-gold, minimum width = 5.9cm, minimum height = .5cm] {};
\end{tikzpicture}

\begin{tikzpicture}
    \node (B) at (-2,0)  {};
    \node (B) at (-1,0)  {$\mathcal{M}_1$};
    \node (B) at (0,0)  {$\mathcal{M}_2$};
    \node (B) at (1,0)  {$\mathcal{M}_3$};
    \node (B) at (2,0)  {$\mathcal{M}_4$};
    \node (B) at (3,0)  {$\boldsymbol{\mathcal{M}_5}$};
    \node (B) at (4,0)  {$\mathcal{M}_6$};
    \node (B) at (5,0)  {$\mathcal{M}_7$};
    \node (B) at (6,0)  {$\mathcal{M}_8$};
    \node (B) at (7,0)  {$\mathcal{M}_9$};
    \node (B) at (8,0)  {$\mathcal{M}_{10}$};
    \node (box0) at (-1.5,0) [draw = ut-gold, minimum width = 3.9cm, minimum height = .5cm, anchor = west] {};
    \node (box) at (3,0) [draw = ut-dark, minimum width = 1cm, minimum height = .5cm] {};
    \node (box2) at (6.1,0) [draw = ut-gold, minimum width = 4.9cm, minimum height = .5cm] {};
\end{tikzpicture}

\begin{tikzpicture}
    \node (B) at (-2,0)  {};
    \node (B) at (-1,0)  {$\mathcal{M}_1$};
    \node (B) at (0,0)  {$\mathcal{M}_2$};
    \node (B) at (1,0)  {$\mathcal{M}_3$};
    \node (B) at (2,0)  {$\mathcal{M}_4$};
    \node (B) at (3,0)  {$\mathcal{M}_5$};
    \node (B) at (4,0)  {$\boldsymbol{\mathcal{M}_6}$};
    \node (B) at (5,0)  {$\mathcal{M}_7$};
    \node (B) at (6,0)  {$\mathcal{M}_8$};
    \node (B) at (7,0)  {$\mathcal{M}_9$};
    \node (B) at (8,0)  {$\mathcal{M}_{10}$};
    \node (box0) at (-1.5,0) [draw = ut-gold, minimum width = 4.9cm, minimum height = .5cm, anchor = west] {};
    \node (box) at (4,0) [draw = ut-dark, minimum width = 1cm, minimum height = .5cm] {};
    \node (box2) at (6.6,0) [draw = ut-gold, minimum width = 3.9cm, minimum height = .5cm] {};
\end{tikzpicture}

\begin{tikzpicture}
    \node (B) at (-2,0)  {};
    \node (B) at (-1,0)  {$\mathcal{M}_1$};
    \node (B) at (0,0)  {$\mathcal{M}_2$};
    \node (B) at (1,0)  {$\mathcal{M}_3$};
    \node (B) at (2,0)  {$\mathcal{M}_4$};
    \node (B) at (3,0)  {$\mathcal{M}_5$};
    \node (B) at (4,0)  {$\mathcal{M}_6$};
    \node (B) at (5,0)  {$\boldsymbol{\mathcal{M}_7}$};
    \node (B) at (6,0)  {$\mathcal{M}_8$};
    \node (B) at (7,0)  {$\mathcal{M}_9$};
    \node (B) at (8,0)  {$\mathcal{M}_{10}$};
    \node (box0) at (-1.5,0) [draw = ut-gold, minimum width = 5.9cm, minimum height = .5cm, anchor = west] {};
    \node (box) at (5,0) [draw = ut-dark, minimum width = 1cm, minimum height = .5cm] {};
    \node (box2) at (7.1,0) [draw = ut-gold, minimum width = 2.9cm, minimum height = .5cm] {};
\end{tikzpicture}

\begin{tikzpicture}
    \node (B) at (-2,0)  {};
    \node (B) at (-1,0)  {$\mathcal{M}_1$};
    \node (B) at (0,0)  {$\mathcal{M}_2$};
    \node (B) at (1,0)  {$\mathcal{M}_3$};
    \node (B) at (2,0)  {$\mathcal{M}_4$};
    \node (B) at (3,0)  {$\mathcal{M}_5$};
    \node (B) at (4,0)  {$\mathcal{M}_6$};
    \node (B) at (5,0)  {$\mathcal{M}_7$};
    \node (B) at (6,0)  {$\boldsymbol{\mathcal{M}_8}$};
    \node (B) at (7,0)  {$\mathcal{M}_9$};
    \node (B) at (8,0)  {$\mathcal{M}_{10}$};
    \node (box0) at (-1.5,0) [draw = ut-gold, minimum width = 6.9cm, minimum height = .5cm, anchor = west] {};
    \node (box) at (6,0) [draw = ut-dark, minimum width = 1cm, minimum height = .5cm] {};
    \node (box2) at (7.6,0) [draw = ut-gold, minimum width = 1.9cm, minimum height = .5cm] {};
\end{tikzpicture}

\begin{tikzpicture}
    \node (B) at (-2,0)  {};
    \node (B) at (-1,0)  {$\mathcal{M}_1$};
    \node (B) at (0,0)  {$\mathcal{M}_2$};
    \node (B) at (1,0)  {$\mathcal{M}_3$};
    \node (B) at (2,0)  {$\mathcal{M}_4$};
    \node (B) at (3,0)  {$\mathcal{M}_5$};
    \node (B) at (4,0)  {$\mathcal{M}_6$};
    \node (B) at (5,0)  {$\mathcal{M}_7$};
    \node (B) at (6,0)  {$\mathcal{M}_8$};
    \node (B) at (7,0)  {$\boldsymbol{\mathcal{M}_9}$};
    \node (B) at (8,0)  {$\mathcal{M}_{10}$};
    \node (box0) at (-1.5,0) [draw = ut-gold, minimum width = 7.9cm, minimum height = .5cm, anchor = west] {};
    \node (box) at (7,0) [draw = ut-dark, minimum width = 1cm, minimum height = .5cm] {};
    \node (box2) at (8.1,0) [draw = ut-gold, minimum width = 1cm, minimum height = .5cm] {};
\end{tikzpicture}

\begin{tikzpicture}
    \node (B) at (-2,0)  {};
    \node (B) at (-1,0)  {$\mathcal{M}_1$};
    \node (B) at (0,0)  {$\mathcal{M}_2$};
    \node (B) at (1,0)  {$\mathcal{M}_3$};
    \node (B) at (2,0)  {$\mathcal{M}_4$};
    \node (B) at (3,0)  {$\mathcal{M}_5$};
    \node (B) at (4,0)  {$\mathcal{M}_6$};
    \node (B) at (5,0)  {$\mathcal{M}_7$};
    \node (B) at (6,0)  {$\mathcal{M}_8$};
    \node (B) at (7,0)  {$\mathcal{M}_9$};
    \node (B) at (8,0)  {$\boldsymbol{\mathcal{M}_{10}}$};
    \node (box0) at (-1.5,0) [draw = ut-gold, minimum width = 8.9cm, minimum height = .5cm, anchor = west] {};
    \node (box) at (8.6,0) [draw = ut-dark, minimum width = 1.1cm, minimum height = .5cm, anchor = east] {};
\end{tikzpicture}
    \caption{Illustration of cross-fitting when splitting by time into folds with adjacent periods. Data is split by time into $K$ folds (here, $K = 10$). Each $\mathcal{M}$ is one fold of data. We train the two ML models on the folds within the golden box. We make predictions with these models and estimate the effects on the fold printed in bold within the grey box.}
    \label{fig:split_time_folds}
\end{figure}

\begin{figure}[ht!]
    \centering
    \begin{tikzpicture}
    \node (B) at (-2,0)  {};
    \node (B) at (-1,0)  {$\boldsymbol{\mathcal{M}_1}$};
    \node (B) at (0,0)  {$\mathcal{M}_2$};
    \node (B) at (1,0)  {$\mathcal{M}_3$};
    \node (B) at (2,0)  {$\mathcal{M}_4$};
    \node (B) at (3,0)  {$\mathcal{M}_5$};
    \node (B) at (4,0)  {$\mathcal{M}_6$};
    \node (B) at (5,0)  {$\mathcal{M}_7$};
    \node (B) at (6,0)  {$\mathcal{M}_8$};
    \node (B) at (7,0)  {$\mathcal{M}_9$};
    \node (B) at (8,0)  {$\mathcal{M}_{10}$};
    \node (box) at (-1.5,0) [draw = ut-dark, minimum width = 2cm, minimum height = .5cm, anchor = west] {};
    \node (box2) at (4.55,0) [draw = ut-gold, minimum width = 8cm, minimum height = .5cm] {};
\end{tikzpicture}

\begin{tikzpicture}
    \node (B) at (-2,0)  {};
    \node (B) at (-1,0)  {$\mathcal{M}_1$};
    \node (B) at (0,0)  {$\boldsymbol{\mathcal{M}_2}$};
    \node (B) at (1,0)  {$\mathcal{M}_3$};
    \node (B) at (2,0)  {$\mathcal{M}_4$};
    \node (B) at (3,0)  {$\mathcal{M}_5$};
    \node (B) at (4,0)  {$\mathcal{M}_6$};
    \node (B) at (5,0)  {$\mathcal{M}_7$};
    \node (B) at (6,0)  {$\mathcal{M}_8$};
    \node (B) at (7,0)  {$\mathcal{M}_9$};
    \node (B) at (8,0)  {$\mathcal{M}_{10}$};
    \node (box) at (0,0) [draw = ut-dark, minimum width = 3cm, minimum height = .5cm] {};
    \node (box2) at (5.05,0) [draw = ut-gold, minimum width = 7cm, minimum height = .5cm] {};
\end{tikzpicture}

\begin{tikzpicture}
    \node (B) at (-2,0)  {};
    \node (B) at (-1,0)  {$\mathcal{M}_1$};
    \node (B) at (0,0)  {$\mathcal{M}_2$};
    \node (B) at (1,0)  {$\boldsymbol{\mathcal{M}_3}$};
    \node (B) at (2,0)  {$\mathcal{M}_4$};
    \node (B) at (3,0)  {$\mathcal{M}_5$};
    \node (B) at (4,0)  {$\mathcal{M}_6$};
    \node (B) at (5,0)  {$\mathcal{M}_7$};
    \node (B) at (6,0)  {$\mathcal{M}_8$};
    \node (B) at (7,0)  {$\mathcal{M}_9$};
    \node (B) at (8,0)  {$\mathcal{M}_{10}$};
    \node (box0) at (-1.5,0) [draw = ut-gold, minimum width = .9cm, minimum height = .5cm, anchor = west] {};
    \node (box) at (1,0) [draw = ut-dark, minimum width = 3cm, minimum height = .5cm] {};
    \node (box2) at (5.55,0) [draw = ut-gold, minimum width = 6cm, minimum height = .5cm] {};
\end{tikzpicture}

\begin{tikzpicture}
    \node (B) at (-2,0)  {};
    \node (B) at (-1,0)  {$\mathcal{M}_1$};
    \node (B) at (0,0)  {$\mathcal{M}_2$};
    \node (B) at (1,0)  {$\mathcal{M}_3$};
    \node (B) at (2,0)  {$\boldsymbol{\mathcal{M}_4}$};
    \node (B) at (3,0)  {$\mathcal{M}_5$};
    \node (B) at (4,0)  {$\mathcal{M}_6$};
    \node (B) at (5,0)  {$\mathcal{M}_7$};
    \node (B) at (6,0)  {$\mathcal{M}_8$};
    \node (B) at (7,0)  {$\mathcal{M}_9$};
    \node (B) at (8,0)  {$\mathcal{M}_{10}$};
    \node (box0) at (-1.5,0) [draw = ut-gold, minimum width = 1.9cm, minimum height = .5cm, anchor = west] {};
    \node (box) at (2,0) [draw = ut-dark, minimum width = 3cm, minimum height = .5cm] {};
    \node (box2) at (6.05,0) [draw = ut-gold, minimum width = 5cm, minimum height = .5cm] {};
\end{tikzpicture}

\begin{tikzpicture}
    \node (B) at (-2,0)  {};
    \node (B) at (-1,0)  {$\mathcal{M}_1$};
    \node (B) at (0,0)  {$\mathcal{M}_2$};
    \node (B) at (1,0)  {$\mathcal{M}_3$};
    \node (B) at (2,0)  {$\mathcal{M}_4$};
    \node (B) at (3,0)  {$\boldsymbol{\mathcal{M}_5}$};
    \node (B) at (4,0)  {$\mathcal{M}_6$};
    \node (B) at (5,0)  {$\mathcal{M}_7$};
    \node (B) at (6,0)  {$\mathcal{M}_8$};
    \node (B) at (7,0)  {$\mathcal{M}_9$};
    \node (B) at (8,0)  {$\mathcal{M}_{10}$};
    \node (box0) at (-1.5,0) [draw = ut-gold, minimum width = 2.9cm, minimum height = .5cm, anchor = west] {};
    \node (box) at (3,0) [draw = ut-dark, minimum width = 3cm, minimum height = .5cm] {};
    \node (box2) at (6.55,0) [draw = ut-gold, minimum width = 4cm, minimum height = .5cm] {};
\end{tikzpicture}

\begin{tikzpicture}
    \node (B) at (-2,0)  {};
    \node (B) at (-1,0)  {$\mathcal{M}_1$};
    \node (B) at (0,0)  {$\mathcal{M}_2$};
    \node (B) at (1,0)  {$\mathcal{M}_3$};
    \node (B) at (2,0)  {$\mathcal{M}_4$};
    \node (B) at (3,0)  {$\mathcal{M}_5$};
    \node (B) at (4,0)  {$\boldsymbol{\mathcal{M}_6}$};
    \node (B) at (5,0)  {$\mathcal{M}_7$};
    \node (B) at (6,0)  {$\mathcal{M}_8$};
    \node (B) at (7,0)  {$\mathcal{M}_9$};
    \node (B) at (8,0)  {$\mathcal{M}_{10}$};
    \node (box0) at (-1.5,0) [draw = ut-gold, minimum width = 3.9cm, minimum height = .5cm, anchor = west] {};
    \node (box) at (4,0) [draw = ut-dark, minimum width = 3cm, minimum height = .5cm] {};
    \node (box2) at (7.05,0) [draw = ut-gold, minimum width = 3cm, minimum height = .5cm] {};
\end{tikzpicture}

\begin{tikzpicture}
    \node (B) at (-2,0)  {};
    \node (B) at (-1,0)  {$\mathcal{M}_1$};
    \node (B) at (0,0)  {$\mathcal{M}_2$};
    \node (B) at (1,0)  {$\mathcal{M}_3$};
    \node (B) at (2,0)  {$\mathcal{M}_4$};
    \node (B) at (3,0)  {$\mathcal{M}_5$};
    \node (B) at (4,0)  {$\mathcal{M}_6$};
    \node (B) at (5,0)  {$\boldsymbol{\mathcal{M}_7}$};
    \node (B) at (6,0)  {$\mathcal{M}_8$};
    \node (B) at (7,0)  {$\mathcal{M}_9$};
    \node (B) at (8,0)  {$\mathcal{M}_{10}$};
    \node (box0) at (-1.5,0) [draw = ut-gold, minimum width = 4.9cm, minimum height = .5cm, anchor = west] {};
    \node (box) at (5,0) [draw = ut-dark, minimum width = 3cm, minimum height = .5cm] {};
    \node (box2) at (7.55,0) [draw = ut-gold, minimum width = 2cm, minimum height = .5cm] {};
\end{tikzpicture}

\begin{tikzpicture}
    \node (B) at (-2,0)  {};
    \node (B) at (-1,0)  {$\mathcal{M}_1$};
    \node (B) at (0,0)  {$\mathcal{M}_2$};
    \node (B) at (1,0)  {$\mathcal{M}_3$};
    \node (B) at (2,0)  {$\mathcal{M}_4$};
    \node (B) at (3,0)  {$\mathcal{M}_5$};
    \node (B) at (4,0)  {$\mathcal{M}_6$};
    \node (B) at (5,0)  {$\mathcal{M}_7$};
    \node (B) at (6,0)  {$\boldsymbol{\mathcal{M}_8}$};
    \node (B) at (7,0)  {$\mathcal{M}_9$};
    \node (B) at (8,0)  {$\mathcal{M}_{10}$};
    \node (box0) at (-1.5,0) [draw = ut-gold, minimum width = 5.9cm, minimum height = .5cm, anchor = west] {};
    \node (box) at (6,0) [draw = ut-dark, minimum width = 3cm, minimum height = .5cm] {};
    \node (box2) at (8.05,0) [draw = ut-gold, minimum width = 1cm, minimum height = .5cm] {};
\end{tikzpicture}

\begin{tikzpicture}
    \node (B) at (-2,0)  {};
    \node (B) at (-1,0)  {$\mathcal{M}_1$};
    \node (B) at (0,0)  {$\mathcal{M}_2$};
    \node (B) at (1,0)  {$\mathcal{M}_3$};
    \node (B) at (2,0)  {$\mathcal{M}_4$};
    \node (B) at (3,0)  {$\mathcal{M}_5$};
    \node (B) at (4,0)  {$\mathcal{M}_6$};
    \node (B) at (5,0)  {$\mathcal{M}_7$};
    \node (B) at (6,0)  {$\mathcal{M}_8$};
    \node (B) at (7,0)  {$\boldsymbol{\mathcal{M}_9}$};
    \node (B) at (8,0)  {$\mathcal{M}_{10}$};
    \node (box0) at (-1.5,0) [draw = ut-gold, minimum width = 6.9cm, minimum height = .5cm, anchor = west] {};
    \node (box) at (7,0) [draw = ut-dark, minimum width = 3.05cm, minimum height = .5cm] {};
\end{tikzpicture}

\begin{tikzpicture}
    \node (B) at (-2,0)  {};
    \node (B) at (-1,0)  {$\mathcal{M}_1$};
    \node (B) at (0,0)  {$\mathcal{M}_2$};
    \node (B) at (1,0)  {$\mathcal{M}_3$};
    \node (B) at (2,0)  {$\mathcal{M}_4$};
    \node (B) at (3,0)  {$\mathcal{M}_5$};
    \node (B) at (4,0)  {$\mathcal{M}_6$};
    \node (B) at (5,0)  {$\mathcal{M}_7$};
    \node (B) at (6,0)  {$\mathcal{M}_8$};
    \node (B) at (7,0)  {$\mathcal{M}_9$};
    \node (B) at (8,0)  {$\boldsymbol{\mathcal{M}_{10}}$};
    \node (box0) at (-1.5,0) [draw = ut-gold, minimum width = 7.9cm, minimum height = .5cm, anchor = west] {};
    \node (box) at (8.55,0) [draw = ut-dark, minimum width = 2.05cm, minimum height = .5cm, anchor = east] {};
\end{tikzpicture}
    \caption{Illustration of ``neighbors-left-out cross-fitting" \citep{semenova_inference_2023}. Data is split by time into $K$ folds (here, $K = 10$). Each $\mathcal{M}$ is one fold of data. We train the two ML models on the folds within the golden box. We make predictions with these models and estimate the effects on the fold printed in bold. The other folds within the grey box are in the immediate neighborhood of the bold fold and excluded from both training and estimation.}
    \label{fig:split_NLO}
\end{figure}

\clearpage

\FloatBarrier

\subsubsection{Results for cross-fitting with different \texorpdfstring{$N$}{N} and \texorpdfstring{$T$}{T}} \label{apx:cf_resNT}


\begin{figure}[ht!]
    \centering
    \includegraphics[width=\textwidth]{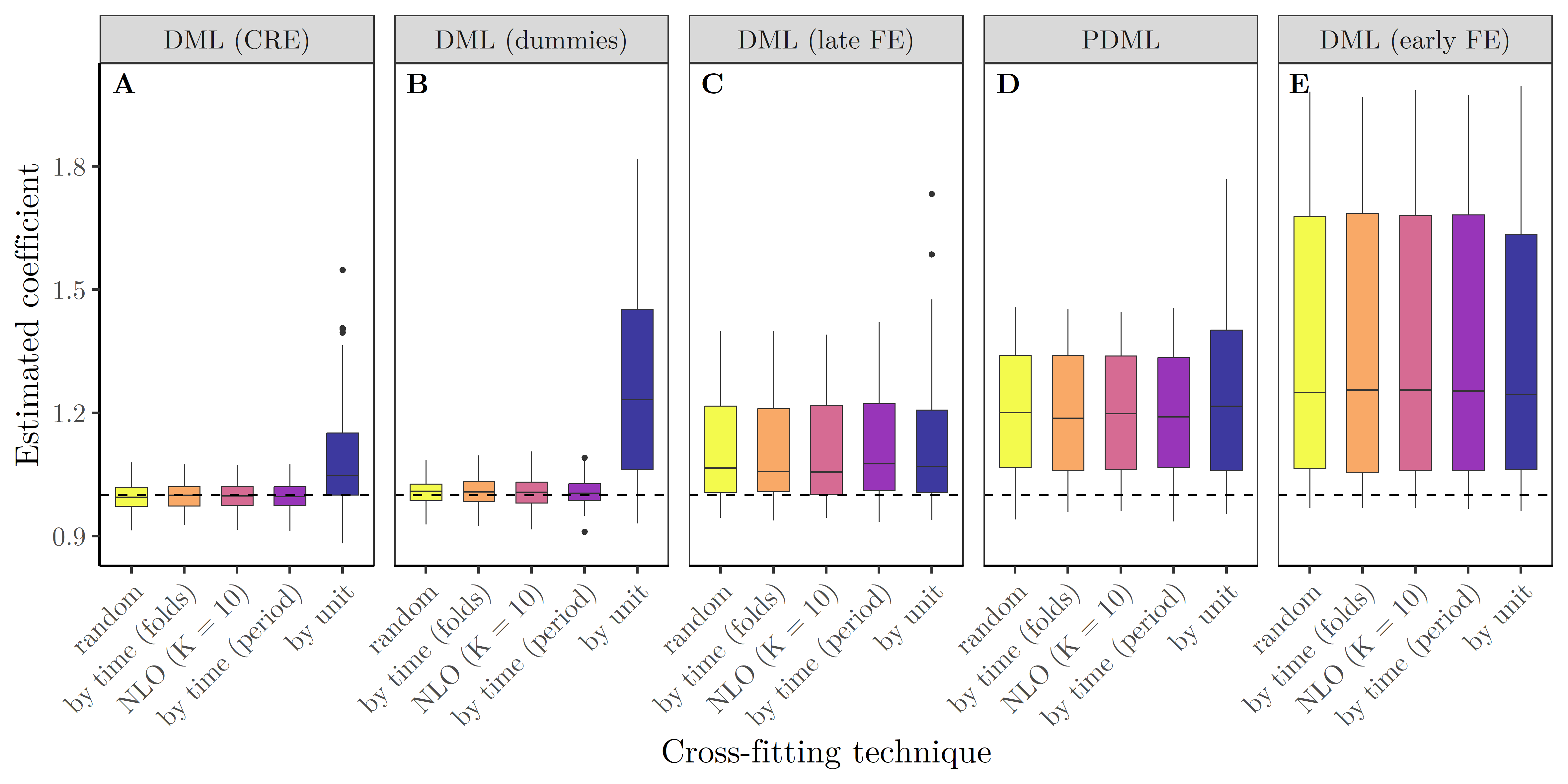}
    \caption{\label{fig:split_resN10_T500}Results for utilizing different cross-fitting techniques (Table \ref{tab:split_DML_approaches}) within various DML estimators for \underline{$N = 10$} and \underline{$T = 500$}. The vertical axis depicts the estimated coefficient. The dashed line marks the true causal effect ($\beta = 1$). The boxplots show the distribution of estimated coefficients across 100 simulated datasets for each method.  Data is generated according to DGP (3), with one observed confounder, u-shaped functional forms and a large degree of autocorrelation ($\rho = .9$). NLO: neighbors-left-out cross-fitting.}
\end{figure}



\begin{figure}[ht!]
    \centering
    \includegraphics[width=\textwidth]{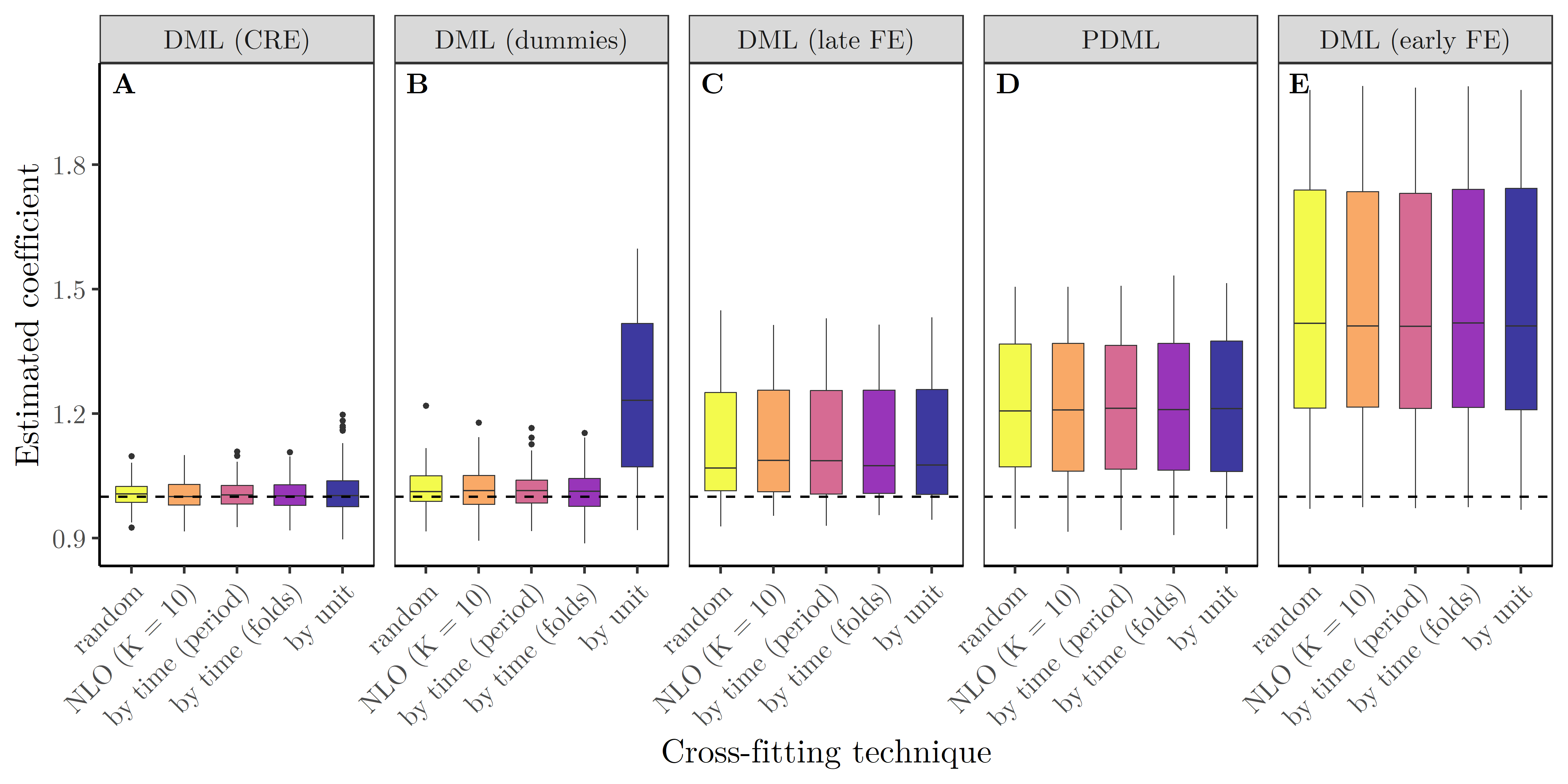}
    \caption{\label{fig:split_resN250_T20}Results for utilizing different cross-fitting techniques (Table \ref{tab:split_DML_approaches}) within various DML estimators  for \underline{$N = 250$} and \underline{$T = 20$}. The vertical axis depicts the estimated coefficient. The dashed line marks the true causal effect ($\beta = 1$). The boxplots show the distribution of estimated coefficients across 100 simulated datasets for each method.  Data is generated according to DGP (3), with one observed confounder, u-shaped functional forms and a large degree of autocorrelation ($\rho = .9$). NLO: neighbors-left-out cross-fitting.}
\end{figure}

\FloatBarrier

\clearpage

\subsection{Computational efficiency of different approaches} \label{sec:comp_eff}

In addition to their varying estimation performance, the different considered estimation approaches also strongly differ in their computational efficiency (Table \ref{tab:comp_eff}). Most notably, DML with dummies is computationally much more expensive than the alternatives, especially in settings where the number of units and thus dummy variables gets large. In our baseline setting, the estimation for a single dataset with $N = 500$ units and $T = 10$ periods, using 5-fold cross-fitting, is approximately 42.8 times slower compared to the second slowest method (DML with CRE). This effect is less pronounced in settings with fewer units (and thus fixed effects). 

\begin{table}[ht!]
\caption{Computation times (in seconds) for different approaches and different data structures}
\label{tab:comp_eff}
\centering
\small
\resizebox{\linewidth}{!}{
\begin{tabular}[t]{p{.2\textwidth}>{\raggedleft}p{.2\textwidth}>{\raggedleft}p{.2\textwidth}>{\raggedleft}p{.2\textwidth}>{\raggedleft\arraybackslash}p{.2\textwidth}}
\toprule
\textbf{Method} & \textbf{$N=500$ / $T=10$} & \textbf{$N=100$ / $T=50$} & \textbf{$N=50$ / $T=100$} & \textbf{$N=10$ / $T=500$}\\
\midrule
DML (early FE) & 4.56 & 5.10 & 4.54 & 5.61\\
PDML & 6.38 & 6.78 & 6.56 & 6.72\\
DML (late FE) & 6.40 & 6.71 & 6.63 & 6.89\\
DML (CRE) & 7.69 & 8.22 & 7.79 & 7.89\\
DML (dummies) & 329.43 & 47.67 & 23.56 & 13.09\\
\bottomrule
\multicolumn{5}{p{1.1\textwidth}}{\footnotesize\textit{Note: }$N$: number of units, $T$: number of periods. Reported execution times are the averages of five iterations for each method-dataset combination. We simulated the data according to DGP (3) of the baseline simulation setting. For this simple comparison, we computed each method on a standard laptop with an Intel Core i5-8365U 4-core CPU with 1.60 GHz and 16 GB of RAM.}
\end{tabular}}
\end{table}

\FloatBarrier

\clearpage

\subsection{Further settings and results}

\subsubsection{Results for intermediate numbers of \texorpdfstring{$N$}{N} and \texorpdfstring{$T$}{T}} \label{apx:furth_resNT}

\begin{figure}[ht!]
    \centering
    \includegraphics[width=\textwidth]{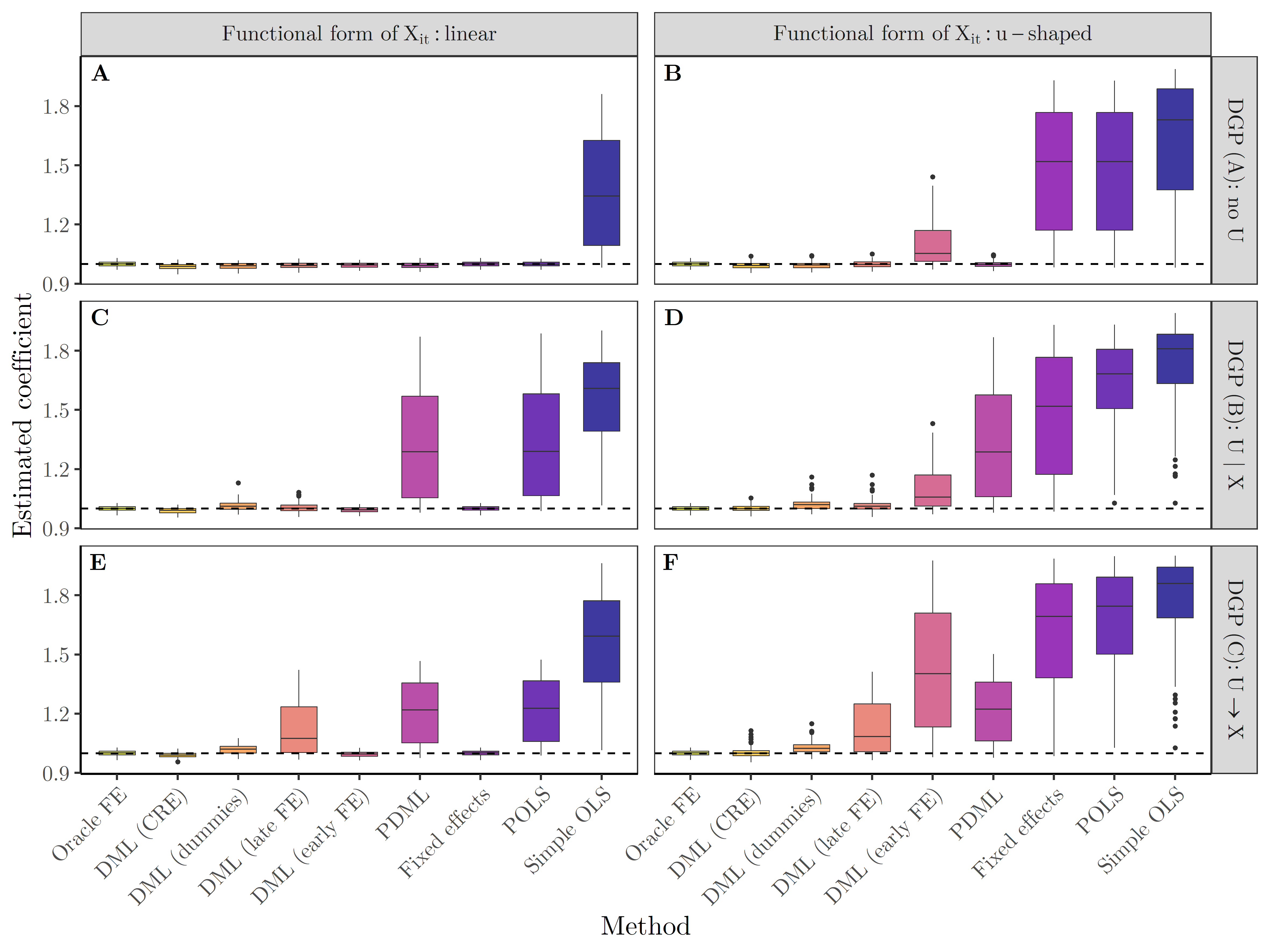}
    \caption{\label{fig:n100t50}Results for the setting with \underline{$N=100$} units and \underline{$T = 50$} periods. The horizontal axis displays the different methods from Table \ref{tab:methods_impl}. The vertical axis depicts the estimated coefficient. The dashed line marks the true causal effect ($\beta = 1$). The boxplots show the distribution of estimated coefficients across 100 simulated datasets for each method. 
    The three rows contain three different DGPs: ``$no\, U$" indicates no unobserved heterogeneity, ``$U\, |\, X$" means the unobserved heterogeneity influences treatment and outcome, but not confounders, and ``$U \rightarrow X$" means the unobserved heterogeneity also influences the confounders.}
\end{figure}

\begin{figure}[ht!]
    \centering
    \includegraphics[width=\textwidth]{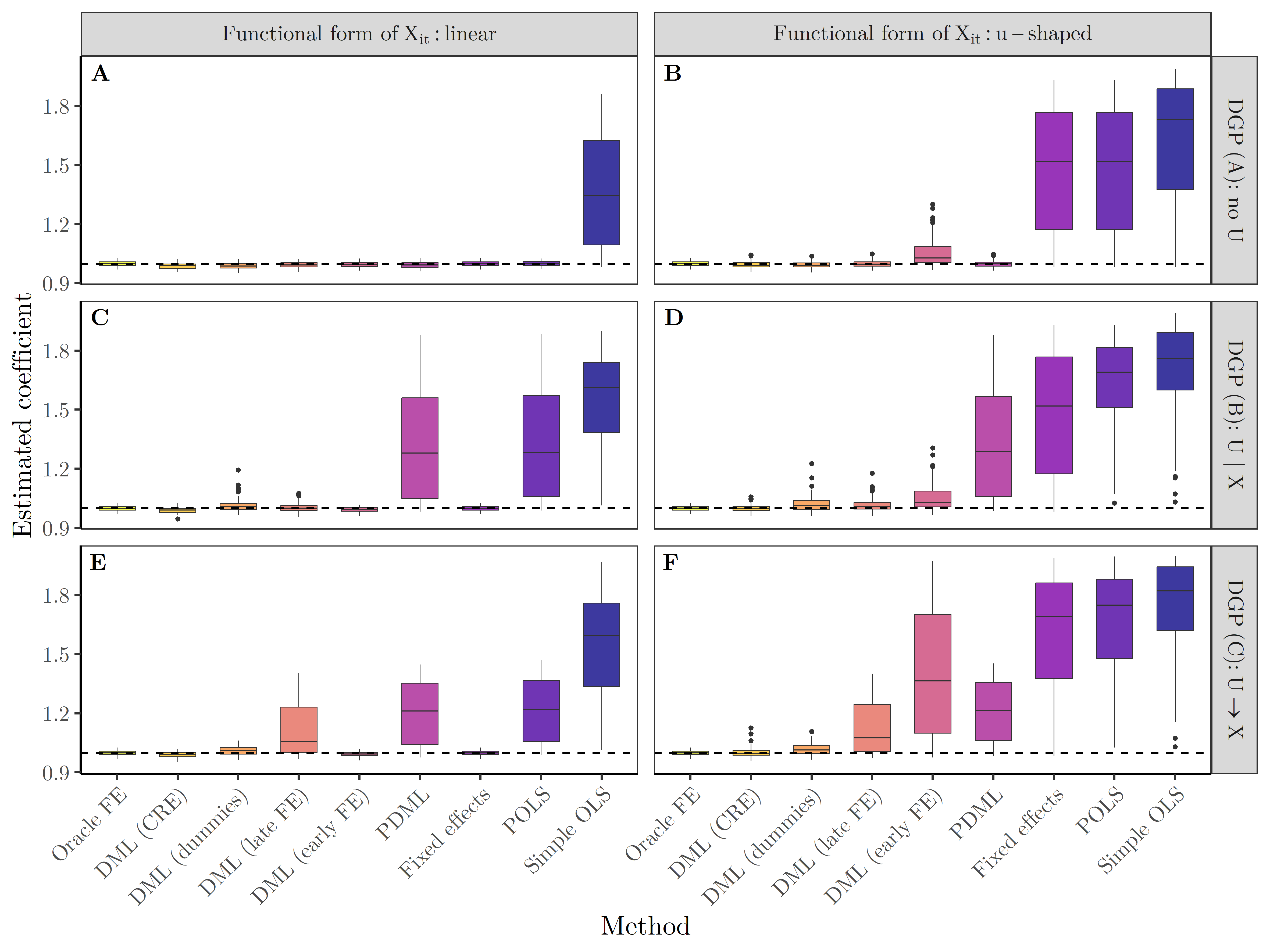}
    \caption{\label{fig:n50t100}Results for the setting with \underline{$N=50$} units and \underline{$T = 100$} periods. The horizontal axis displays the different methods from Table \ref{tab:methods_impl}. The vertical axis depicts the estimated coefficient. The dashed line marks the true causal effect ($\beta = 1$). The boxplots show the distribution of estimated coefficients across 100 simulated datasets for each method. 
    The three rows contain three different DGPs: ``$no\, U$" indicates no unobserved heterogeneity, ``$U\, |\, X$" means the unobserved heterogeneity influences treatment and outcome, but not confounders, and ``$U \rightarrow X$" means the unobserved heterogeneity also influences the confounders.}
\end{figure}

\FloatBarrier

\clearpage

\subsubsection{Varying the number of observed confounders in linear settings or with more periods} \label{apx:furth_nconf_lin_per}

\begin{figure}[ht!]
    \centering
    \includegraphics[width=\textwidth]{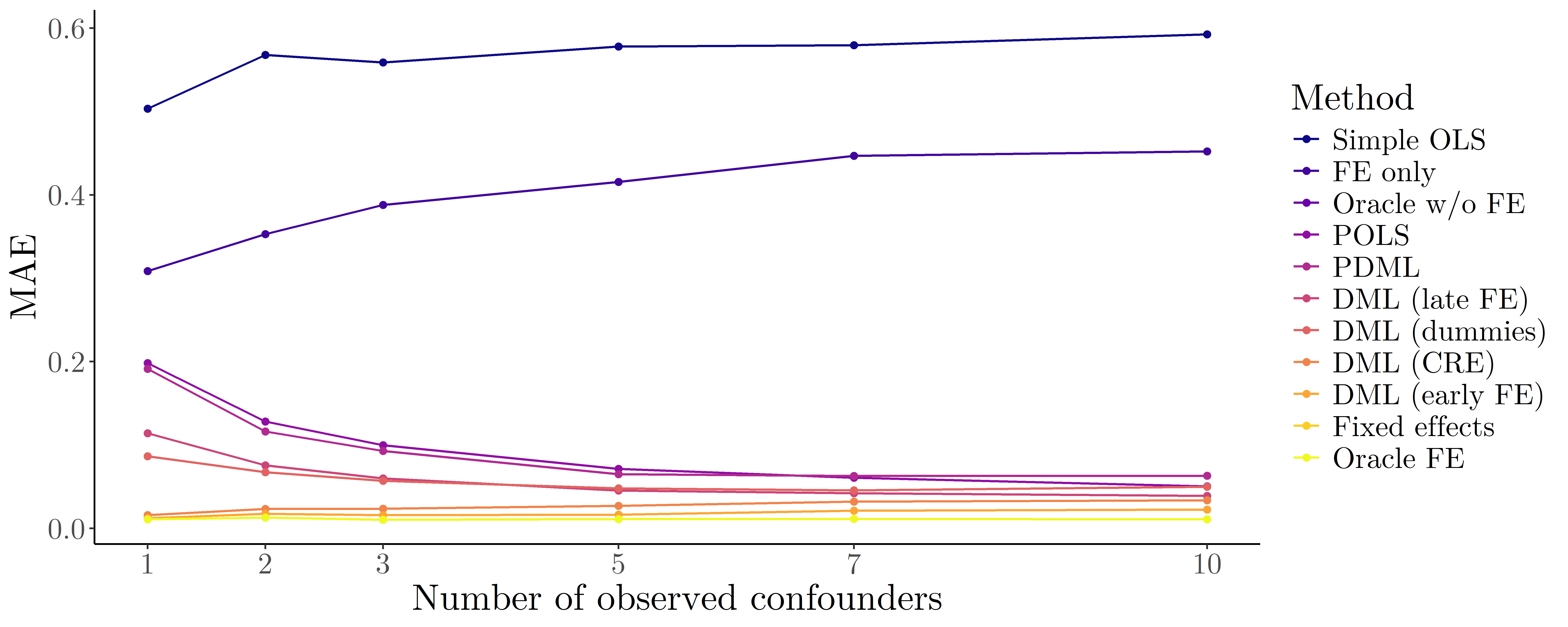}
    \caption{\label{fig:nconf_linear}Mean absolute error in the estimated coefficient across 100 simulations by number of observed confounders. The simulated confounding influence is \underline{linear}, the causal structure is (C), i.e., $U_i \rightarrow X_{it}$. $N = 500$, $T = 10$.}
\end{figure}

\begin{figure}[ht!]
    \centering
    \includegraphics[width=\textwidth]{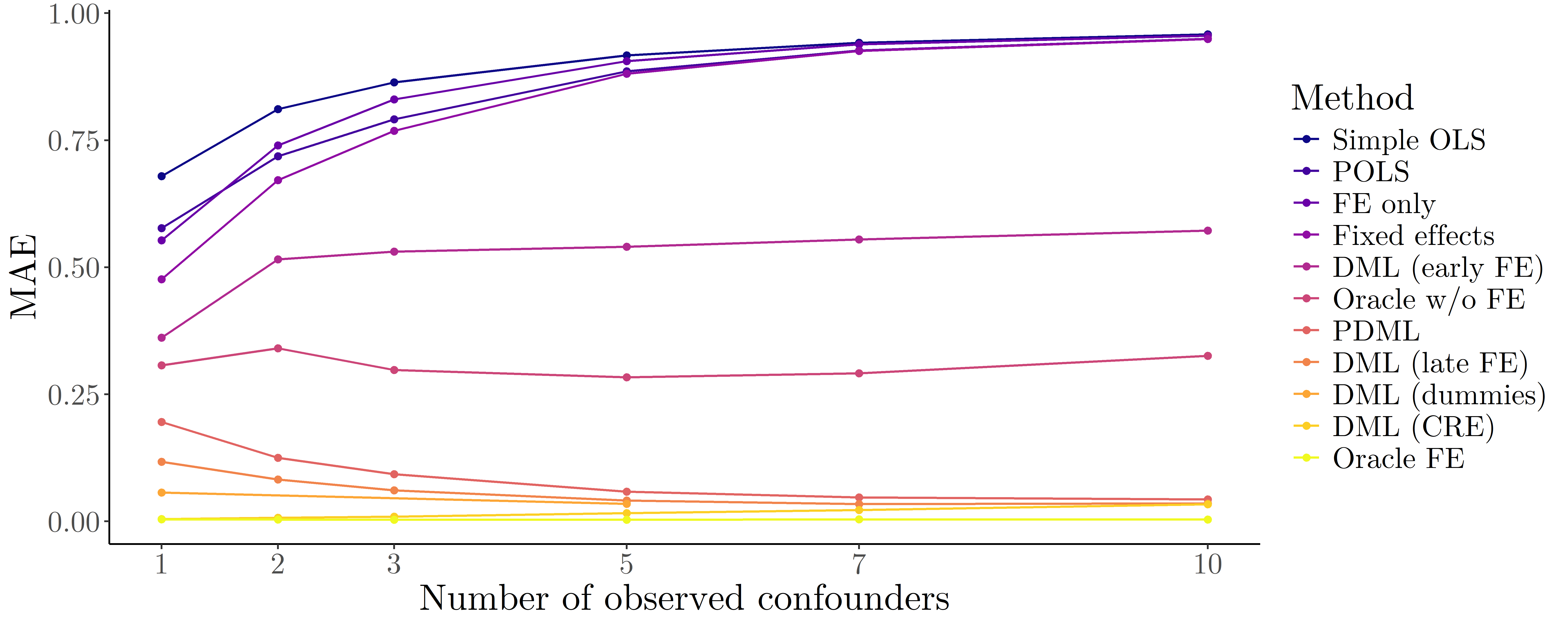}
    \caption{\label{fig:nconf500_100}Mean absolute error in the estimated coefficient across 100 simulations by number of observed confounders. The simulated confounding influence is u-shaped, the causal structure is (C), i.e., $U_i \rightarrow X_{it}$. $N = 500$, \underline{$T = 100$}. We computed DML (dummies) only for 1 and 5 confounders due to unreasonably high computational costs for this data size.}
\end{figure}

\FloatBarrier
\clearpage

\subsubsection{Increasing sample size by increasing the number of periods} \label{apx:furth_byperiod}

\begin{figure}[ht!]
    \centering
    \includegraphics[width=\textwidth]{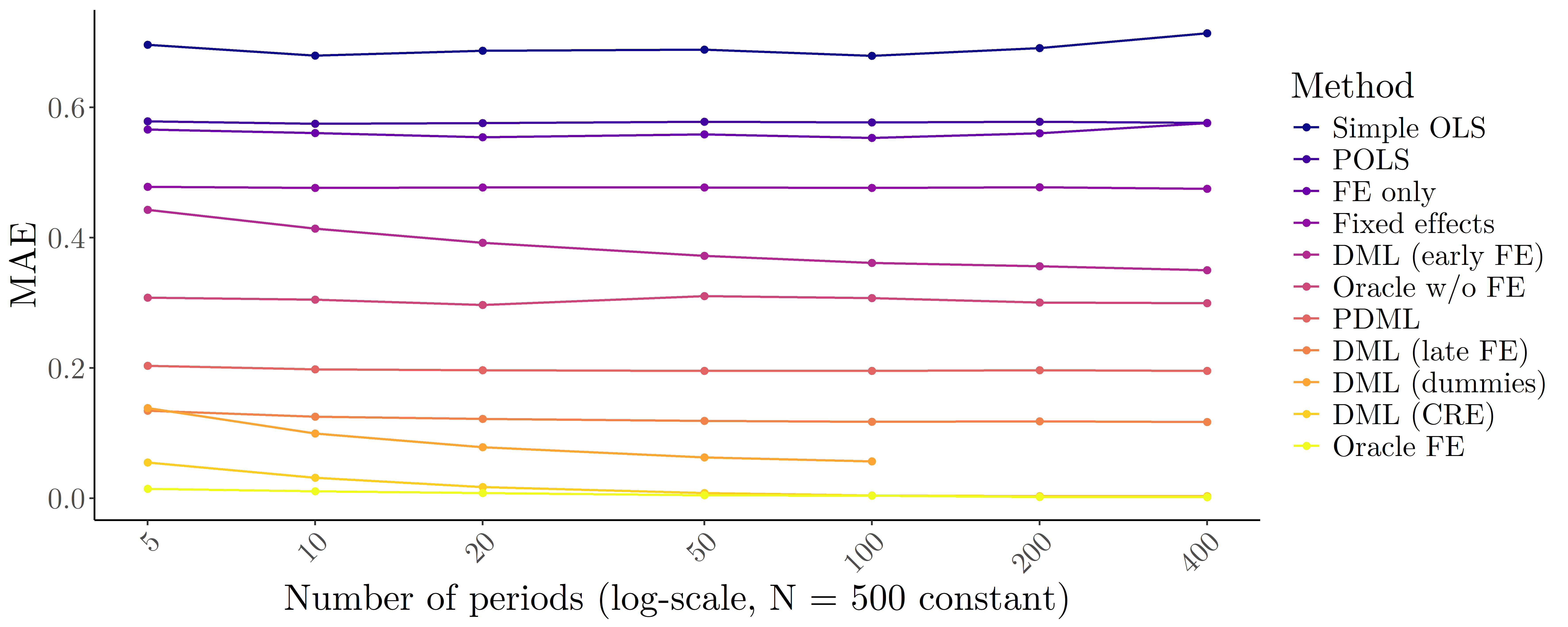}
    \caption{\label{fig:ss1T_nconf}Mean absolute error in the estimated coefficient across 100 simulations for \textbf{1} observed confounder by the number of periods. The number of units is fixed at $N = 500$. The simulated confounding influence is u-shaped, the causal structure is (C), i.e., $U_i \rightarrow X_{it}$. We computed DML (dummies) only for up to $T = 100$, as it becomes computationally too costly for larger values.}
\end{figure}

\begin{figure}[ht!]
    \centering
    \includegraphics[width=\textwidth]{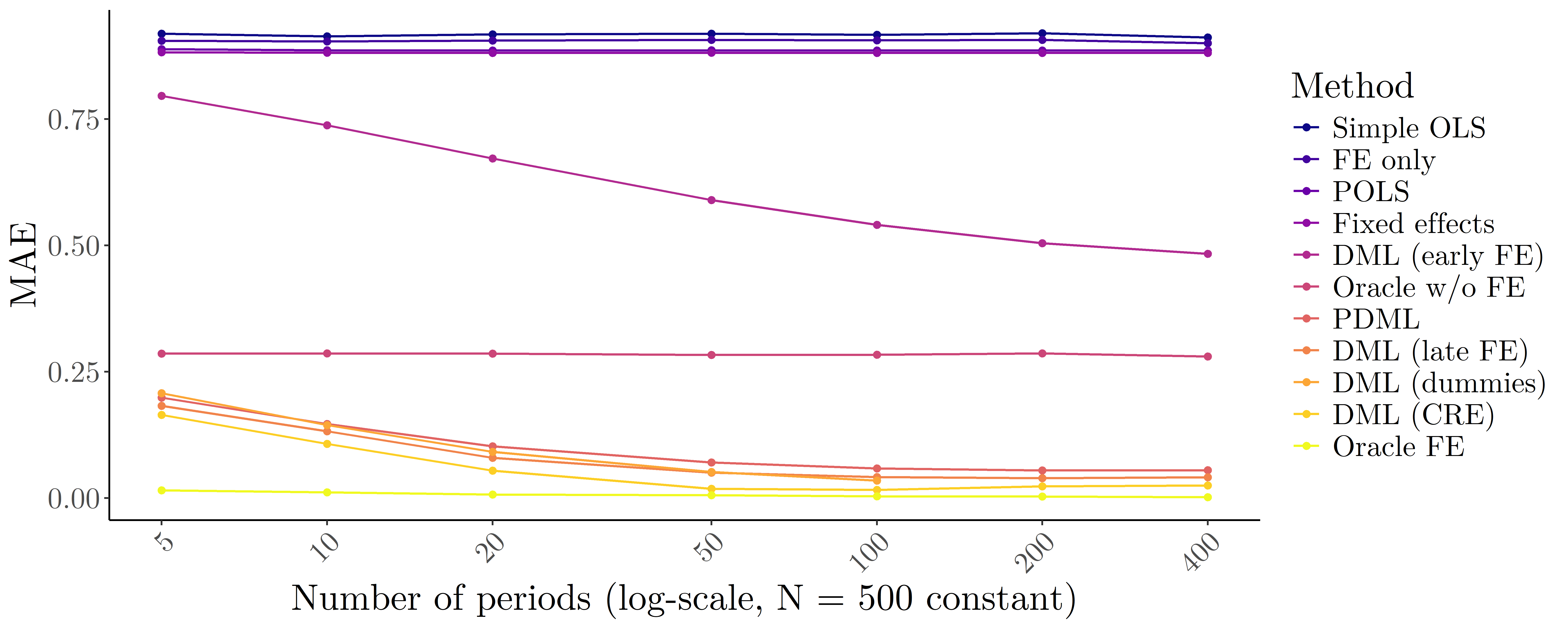}
    \caption{\label{fig:ss5T_nconf}Mean absolute error in the estimated coefficient across 100 simulations for \textbf{5} observed confounders by the number of periods. The number of units is fixed at $N = 500$. The simulated confounding influence is u-shaped, the causal structure is (C), i.e., $U_i \rightarrow X_{it}$. We computed DML (dummies) only for up to $T = 100$, as it becomes computationally too costly for larger values.}
\end{figure}

\FloatBarrier

\clearpage

\end{document}